\documentstyle[aps,prd,amstex,epsfig]{revtex}

\begin{document}

\author{Christophe Ringeval} \address{Institut d'Astrophysique
de Paris, 98bis boulevard Arago, 75014 Paris, France.}

\title{Fermionic massive modes along cosmic strings}
\date{\today}

\maketitle

\newcommand{\ETAL}{{\it et al.}}
\newcommand{\ud}{{\mathrm{d}}}
\newcommand{\ue}{{\mathrm{e}}}
\newcommand{\uV}{{\mathrm{V}}}
\newcommand{\Chi}{{\mathcal{X}}}
\newcommand{\Ferm}{{\mathcal{F}}}
\newcommand{\Rho}{{\mathcal{R}}}
\newcommand{\norm}{{\mathcal{N}}}
\newcommand{\M}{{\mathcal{M}}}
\newcommand{\Pstate}{{\mathcal{P}}}
\newcommand{\Eferm}{{\mathcal{E}}}
\newcommand{\Rhop}{\widehat{{\mathcal{R}}}}
\newcommand{\cpsir}{c_{\psi_{\mathrm{R}}}}
\newcommand{\cpsil}{c_{\psi_{\mathrm{L}}}}
\newcommand{\cchir}{c_{\chi_{\mathrm{R}}}}
\newcommand{\cchil}{c_{\chi_{\mathrm{L}}}}
\newcommand{\cphi}{c_{\phi}}
\newcommand{\chip}{{\chi_{\mathcal{P}}}}
\newcommand{\psip}{{\psi_{\mathcal{P}}}}
\newcommand{\cfl}{c_{\Ferm_{\mathrm{L}}}}
\newcommand{\cfr}{c_{\Ferm_{\mathrm{R}}}}
\newcommand{\jbt}{\tilde{\overline{j}}}
\newcommand{\I}{I}
\newcommand{\miv}{\overline{m}}
\newcommand{\xit}{\widetilde{\xi}}
\newcommand{\rhot}{\widetilde{\rho}}
\newcommand{\zetat}{\widetilde{\zeta}}
\newcommand{\at}{\widetilde{\alpha}}
\newcommand{\ab}{\overline{\alpha}}
\newcommand{\ah}{\widehat{\alpha}}
\newcommand{\bt}{\widetilde{\beta}}
\newcommand{\nut}{\widetilde{\nu}}
\newcommand{\nub}{\overline{\nu}}
\newcommand{\Sigmab}{\overline{\Sigma}}
\newcommand{\ft}{\widetilde{f}}
\newcommand{\gt}{\widetilde{g}}
\newcommand{\xih}{\widehat{\xi}}
\newcommand{\uh}{\widehat{u}}
\newcommand{\vh}{\widehat{v}}
\newcommand{\xpar}{x_{\parallel}}
\newcommand{\xper}{x_{\perp}}
\newcommand{\syse}{\left({\mathcal{S}}_{\varepsilon}\right)}
\newcommand{\sysp}{\left({\mathcal{S}}_+\right)}
\newcommand{\sysm}{\left({\mathcal{S}}_-\right)}
\newcommand{\vsep}{\vspace{4pt}}

\begin{abstract}
The influence on cosmic string dynamics of fermionic massive bound
states propagating in the vortex, and getting their mass only from
coupling to the string forming Higgs field, is studied. Such massive
fermionic currents are numerically found to exist for a wide range of
model parameters and seen to modify drastically the usual string
dynamics coming from the zero mode currents alone. In particular, by
means of a quantization procedure, a new equation of state describing
cosmic strings with any kind of fermionic current, massive or
massless, is derived and found to involve, at least, one state
parameter per trapped fermion species. This equation of state
exhibits transitions from subsonic to supersonic regimes while
the massive modes are filled.

\end{abstract}


\section{Introduction}

Since it was realized that some early universe phase transitions might
lead to the formation of topological defects~\cite{kibble}, cosmic
strings have been the subject of intense work within the context of
cosmology~\cite{strings}. The large scale structure generated by an
ordinary string network in an expanding universe, as well as its
imprint on the microwave background, have thus been
derived~\cite{CMBR,BPRS} in order to state on their significance in
the wide range of mechanisms in which they had been originally
involved~\cite{zeldov,vilenkin}. These predictions, compared with the
observations therefore constrain the symmetry breaking schemes
effectively realized in the early Universe. These, associated with the
most recent data for the microwave background
anisotropies~\cite{BOOM}, even seem to show that such ordinary string
networks could not have play the dominant role in the Universe
evolution, thereby all the more so constraining the particle physics
symmetries leading to their formation. However, as was recently
shown~\cite{BPRS}, a non-negligible fraction of such defects could
have contributed to the overall cosmic microwave background (CMB)
anisotropies.

Meanwhile, it was shown by Witten~\cite{witten} that in realistic
physical models, involving various couplings of the string forming
Higgs field to other scalar or fermion fields, currents could build
along the strings, turning them into ``superconducting wires.''
Without even introducing couplings with the electromagnetic
fields~\cite{otw}, the breaking of Lorentz invariance along the vortex
induced by such currents may drastically modify the string properties,
and thus, the cosmological evolution of the associated networks. In
particular, cosmic string loops can reach centrifugally supported
equilibrium state, called vortons~\cite{davisRL}, that would
completely dominate the Universe~\cite{brandenberger}. Theories
predicting stable vortons thus turn out to be incompatible with
observational cosmology, hence the particular interest focused on
``superconducting'' models.

Unfortunately, all the new properties and cosmological consequences
stemming from string conductivity have not yet been clearly
established, because of the complicated, and somehow arbitrary,
microphysics possible in these models. However, although the
microscopic properties induced by such currents depend on the explicit
underlying field theory~\cite{massF,vecmass}, a macroscopic formalism
was introduced by Carter~\cite{formal} which permits a unified
description of the string dynamics through the knowledge of its energy
per unit length $U$ and tension $T$. These ones end up being functions
of a so-called state parameter $w$, as the current itself, through an
equation of state. Such a formalism is, in particular, well designed
for scalar currents, as shown in, e.g., Refs.~\cite{neutral,enon0}:
due to their bosonic nature, all trapped scalar particles go into the
lowest accessible state, and thus can be described through the
classical values taken by the relevant scalar fields~\cite{bps}. The
induced gravitational field~\cite{Garrigapeter,peterpuy} or the back
reaction effects~\cite{nospring} depend only on this state
parameter. The classical string
stability~\cite{cartermartin,martinpeter} has already been
investigated for various equations of state relating $U$ and $T$, on
the basis of scalar and chiral currents
microphysics~\cite{cp1,cp2}. Moreover, it was also shown, through a
semiclassical approach, that fermionic current carrying cosmic
strings, even though in principle involving more than one state
parameter~\cite{ringeval}, can also be described by an equation of
state of the so-called ``fixed trace'' kind, i.e., $U+T=2M^2$.  Such a
relationship has the property of allowing stable loop configurations
to exist, at least at the classical
level~\cite{martinpeter}. Nevertheless, these results have been
derived for fermionic currents flowing along the string in the form of
zero modes only, as they were originally introduced by
Witten~\cite{witten}, although it was shown that the fermions may also
be trapped in the vortex with nonvanishing masses~\cite{davisC}: hence
the following work in which the influence of such massive modes is
studied for the simplest of all fermionic Witten model.

In this paper, after deriving numerically the relevant properties of
the trapped massive wave solutions of the Dirac equation in the
vortex, we show that the quantization procedure, originally performed
to deal with the fermionic zero modes~\cite{ringeval}, can be
generalized to include the massive ones, and leads to a new equation
of state with more than one state parameter. In particular, it is
found that the fixed trace equation of state, that holds for
massless fermionic currents alone, is no longer verified. Besides, the
massive modes are actually found to rapidly dominate the string
dynamics, thereby modifying the classical vorton stability induced by
the zero modes alone.

Let us sketch the lines along which this work is made. In
Sec.~\ref{modele}, the model and the notations are set, while we
derive the equations of motion. Then, in Sec.~\ref{modemassif}, by
means of a separation between transverse and longitudinal degrees of
freedom of the spinor fields, the massive wave solutions along the
string are computed numerically for a wide range of fermion charges
and coupling constants. The constraint of transverse normalizability
is found to be satisfied only for particular values of the trapped
modes mass, $\miv$ say, whose dependence with the model parameters is
investigated. The two-dimensional quantization of the $\miv$ normalizable
massive modes is then performed in Sec.~\ref{quantization}, using the
canonical procedure. In the way previously discussed in the
case of zero modes~\cite{ringeval}, the conserved quantities, i.e.,
energy-momentum tensor and charge currents, are then expressed
in their quantum form. Their average
values, in the zero-temperature case, and infinite string limit, lead
to macroscopic expressions for the energy per unit length $U$ and
tension $T$ which end up being functions of the number densities of
fermions propagating along the string. Their derivation and extension
to any kind and number of fermionic carriers is performed in
Sec.~\ref{etat}, while the cosmological consequences of this new
analysis are briefly discussed in the concluding section.

\section{Model}
\label{modele}

We shall consider here an Abelian Higgs model with scalar $\Phi$ and
gauge field $B_\mu$, coupled, following Witten~\cite{witten}, to two
spinor fields, $\Psi$ and $\Chi$ say. Since we are only interested in
the purely dynamical effects the current may induce on the strings,
we will not
consider any additional electromagneticlike coupling of the fermion
fields to an extra gauge field. Thus, we consider here the so-called
``neutral limit''~\cite{neutral}

\subsection{Microscopic Lagrangian}

The previous assumptions imply one needs one local $U(1)$ symmetry
which is spontaneously broken through the Higgs mechanism, yielding
vortices formation. The Higgs field is chosen as complex scalar field
with conserved charge $q \cphi$ under the local $U(1)$ symmetry,
associated with a gauge vector field $B_\mu$. The two spinor fields
acquire masses from a chiral coupling to the Higgs field, and have
opposite electromagnetic charges in order for the full
(four-dimensional) model to be anomaly free~\cite{witten}. Under the
broken symmetry they also have conserved charges $q \cpsir$, $q
\cpsil$ and $q \cchir$, $q \cchil$ for their right- and left-handed
parts, respectively. With ${\mathcal{L}}_{\mathrm{h}}$,
${\mathcal{L}}_{\mathrm{g}}$ and ${\mathcal{L}}_\psi$,
${\mathcal{L}}_\chi$, the Lagrangian in the Higgs, gauge, and
fermionic sectors, respectively, the theory reads
\begin{equation}
\label{lagrangien}
{\mathcal{L}}={\mathcal{L}}_{\mathrm{h}} + {\mathcal{L}}_{\mathrm{g}}
+ {\mathcal{L}}_\psi + {\mathcal{L}}_\chi,
\end{equation}
with
\begin{eqnarray}
{\mathcal{L}}_{\mathrm{h}} & = & \frac{1}{2}
(D_{\mu}\Phi)^{\dag}(D^{\mu}\Phi) - V(\Phi), \\
{\mathcal{L}}_{\mathrm{g}} & = & -\frac{1}{4} H_{\mu \nu} H^{\mu \nu},
\\
\label{psilagrangian}
{\mathcal{L}}_\psi & = & \frac{i}{2} \left[\overline{\Psi}
\gamma^{\mu} D_{\mu} \Psi - (\overline{D_{\mu}\Psi}) \gamma^{\mu}
\Psi \right] -g \overline{\Psi} \frac{1+\gamma_5}{2} \Psi \Phi
-g\overline{\Psi} \frac{1-\gamma_5}{2} \Psi \Phi^\ast,
\\
\label{chilagrangian}
{\mathcal{L}}_\chi & = & \frac{i}{2} \left[\overline{{\Chi}}
\gamma^{\mu} D_{\mu} {\Chi} - (\overline{D_{\mu}{\Chi}})
\gamma^{\mu} {\Chi} \right] -g \overline{{\Chi}}
\frac{1+\gamma_5}{2} {\Chi} \Phi^{\ast} - g \overline{{\Chi}}
\frac{1-\gamma_5}{2} {\Chi} \Phi,
\end{eqnarray}
where the $U(1)$ field strength tensor and the scalar potential
are
\begin{eqnarray}
H_{\mu \nu} & = & \nabla_\mu B_\nu - \nabla_\nu B_\mu,
\\
V(\Phi) & = & \frac{\lambda}{8} (|\Phi|^2 - \eta^2)^2,
\end{eqnarray}
while covariant derivatives involve the field charges through
\begin{eqnarray}
D_{\mu}\Phi & = & \left(\nabla_{\mu} + i q c_{\phi}B_\mu \right) \Phi,
\\
D_{\mu}\Psi & = & \left(\nabla_{\mu} + i q \frac{\cpsir+\cpsil}{2}B_\mu
+i q \frac{\cpsir-\cpsil}{2} \gamma_5 B_\mu \right) \Psi,
\\
D_{\mu}{\Chi} & = & \left(\nabla_{\mu} + i q \frac{\cchir+
\cchil}{2}B_\mu +i q \frac{\cchir-\cchil}{2} \gamma_5 B_\mu \right)
{\Chi},
\end{eqnarray}
and the relation
\begin{equation}
\cpsil-\cpsir=c_\phi=\cchir-\cchil
\end{equation}
should hold in order for the Yukawa terms in ${\mathcal{L}}_\psi$ and
${\mathcal{L}}_\chi$ to be gauge invariant.
\subsection{Basic equations}

This theory admits vortex solutions which are expected to form in the
early universe by means of the Kibble mechanism~\cite{kibble}. A
cosmic string configuration can be chosen to lie along the
$z$ axis, and we will use Nielsen-Olesen solutions of the field
equations~\cite{NO}. In cylindrical coordinates, the string forming
Higgs and gauge fields thus read
\begin{equation}
\begin{array}{ccc}
\Phi  =  \varphi(r) \ue^{i n\theta},
& \quad &
B_\mu  =  B(r) \delta_{\mu \theta},
\end{array}
\end{equation}
where the winding number $n$ is an integer, in order for the Higgs
field to be single valued under rotation around the string.
In such vortex background, the equations of motion in the fermionic
sector, for both spinor fields $\Ferm$ read (here and in various places
throughout this paper, we shall denote by $\Ferm$ an arbitrary
fermion, namely a spinor $\Psi$ or $\Chi$)
\begin{eqnarray}
\label{fermmvt}
i \gamma^\mu \nabla_\mu \Ferm & = & \frac{\partial j^\mu_\Ferm}{\partial
\overline{\Ferm}} B_\mu + M_\Ferm \Ferm
\end{eqnarray}
with the fermionic gauge currents 
\begin{eqnarray}
\label{currents}
j^\mu_\Ferm & = & q \displaystyle{\frac{\cfr+ \cfl}{2}} \overline{\Ferm}
\gamma^\mu \Ferm + q \displaystyle{\frac{\cfr - \cfl}{2}}
\overline{\Ferm} \gamma^\mu \gamma_5 \Ferm,
\end{eqnarray}
and the mass terms
\begin{eqnarray}
\label{psihiggs}
M_\psi & = & g \varphi \cos{n \theta} + i g \varphi \gamma_5 \sin{n \theta},
\\
\label{chihiggs}
M_\chi & = & g \varphi \cos{n \theta} - i g \varphi \gamma_5 \sin{n \theta}.
\end{eqnarray}
Note the fermionic currents have an axial and vectorial component
because of the chiral coupling of the fermions to the Higgs field, as
can be seen through the mass terms $M_\Ferm$ in Eqs.~(\ref{psihiggs})
and (\ref{chihiggs}). Moreover, since the Higgs field vanishes in the
string core while taking nonzero vacuum expectation value, $\eta$
say, outside, the mass term acts as an attractive potential. As a
result, fermionic bound states, with energy between zero and $g \eta$,
are expected to exist and propagate in the string core.

\section{Fermionic bound states}
\label{modemassif}

\subsection{Trapped wave solutions}

Since the string is assumed axially symmetric, it is convenient to
look for trapped solutions of the fermionic equations of motion, by
separating longitudinal and transverse dependencies of the spinor
fields. Using the same notations as in Ref.~\cite{ringeval}, the
two-dimensional plane-wave solutions along the string, for both
fermions, read
\begin{equation}
\label{planeansatz}
\begin{array}{lll}
\Psi_{\mathrm{p}}^{(\varepsilon)}  =  \ue^{\varepsilon i(\omega t-kz)}
\left(\begin{array}{l}
\xi_1(r) \ue^{-im_1 \theta} \\
\xi_2(r) \ue^{-im_2 \theta} \\
\xi_3(r) \ue^{-im_3 \theta} \\
\xi_4(r) \ue^{-im_4 \theta}
\end{array} \right),
& \quad &
{\Chi}_{\mathrm{p}}^{(\varepsilon)}  =  \ue^{\varepsilon i(\omega t-kz)}
\left(\begin{array}{l}
\zeta_1(r) \ue^{-il_1 \theta} \\
\zeta_2(r) \ue^{-il_2 \theta} \\
\zeta_3(r) \ue^{-il_3 \theta} \\
\zeta_4(r) \ue^{-il_4 \theta}
\end{array} \right),
\end{array}
\end{equation}
where $\varepsilon=\pm 1$ labels the positive and negative energy
solutions. Similarly to the Higgs field case, the winding numbers of
the fermions, $m_i$ and $l_i$, are necessary integers.  In order to
simplify the notations, it is more convenient to work with
dimensionless scaled fields and coordinates. With
$m_{\mathrm{h}}=\eta\sqrt{\lambda}$ the mass of the Higgs boson, we
can write
\begin{eqnarray}
\varphi=\eta H,
\quad
Q=n + q \cphi \, B,
\quad \textrm{and} \quad
r=\frac{\varrho}{m_{\mathrm{h}}}.
\end{eqnarray}
In the same way, the spinorial components of the $\Psi$ field are
rescaled as
\begin{equation}
\begin{array}{ccccccc}
\vsep
\displaystyle
\xi_1(\varrho) & = & 
\displaystyle 
\frac{m_{\mathrm{h}}}{\sqrt{2\pi}}\sqrt{\omega+k}\,\at_1(\varrho),
& \quad &
\displaystyle
\xi_2(\varrho) & = & 
\displaystyle
i \frac{m_{\mathrm{h}}}{\sqrt{2\pi}}\sqrt{\omega - k}\,\at_2(\varrho), \\
\vsep
\displaystyle
\xi_3(\varrho) & = & 
\displaystyle
\frac{m_{\mathrm{h}}}{\sqrt{2\pi}}\sqrt{\omega-k}\,\at_3(\varrho),
& \quad &
\displaystyle
\xi_4(\varrho) & = &
\displaystyle
i \frac{m_{\mathrm{h}}}{\sqrt{2\pi}}\sqrt{\omega+k}\,\at_4(\varrho).
\end{array}
\end{equation}
In the chiral representation, and with the metric signature
$(+,-,-,-)$, in terms of these new variables, Eqs.~(\ref{fermmvt}) and
(\ref{planeansatz}) yield, for the $\Psi$ field,
\begin{equation}
\label{systadim}
\begin{array}{ccc}
\vsep \displaystyle \ue^{-i(m_1-1)\theta} \left[\frac{\ud \at_1}{\ud
 \varrho} - \ft_1(\varrho) \at_1(\varrho)\right] & = & \displaystyle
 \varepsilon \frac{\miv}{m_{\mathrm{h}}} \ue^{-im_2\theta}
 \at_2(\varrho) - \frac{m_{\mathrm{f}}}{m_{\mathrm{h}}} H(\varrho)
 \ue^{-i(m_4+n) \theta} \at_4(\varrho), \\ \vsep \displaystyle
 \ue^{-i(m_2+1)\theta} \left[\frac{\ud \at_2}{\ud \varrho} -
 \ft_2(\varrho) \at_2(\varrho)\right] & = & \displaystyle -
 \varepsilon \frac{\miv}{m_{\mathrm{h}}} \ue^{-im_1\theta}
 \at_1(\varrho) + \frac{m_{\mathrm{f}}}{m_{\mathrm{h}}}H(\varrho)
 \ue^{-i(m_3+n)\theta} \at_3(\varrho), \\ \vsep \displaystyle
 \ue^{-i(m_3-1)\theta} \left[\frac{\ud\at_3}{\ud \varrho} -
 \ft_3(\varrho) \at_3(\varrho)\right] & = & \displaystyle -
 \varepsilon \frac{\miv}{m_{\mathrm{h}}} \ue^{-i m_4 \theta}
 \at_4(\varrho) + \frac{m_{\mathrm{f}}}{m_{\mathrm{h}}} H(\varrho)
 \ue^{-i(m_2-n)\theta} \at_2(\varrho), \\ \vsep \displaystyle
 \ue^{-i(m_4+1)\theta} \left[\frac{\ud \at_4}{\ud \varrho} -
 \ft_4(\varrho) \at_4(\varrho) \right] & = & \displaystyle \varepsilon
 \frac{\miv}{m_{\mathrm{h}}} \ue^{-i m_3\theta} \at_3(\varrho) -
 \frac{m_{\mathrm{f}}}{m_{\mathrm{h}}} H(\varrho)
 \ue^{-i(m_1-n)\theta} \at_1(\varrho),
\end{array}
\end{equation}
where $m_{\mathrm{f}}=g \eta$ is the fermion mass in the vacuum in
which the Higgs field takes its vacuum expectation value $\eta$, and
$\miv=\sqrt{\omega^2-k^2}$ is the mass of the trapped mode. The
coupling to the gauge field $B_\mu$ appears through the purely radial
functions $\ft$:
\begin{equation}
\begin{array}{ccccccc}
\vsep
\displaystyle
\ft_1(\varrho) & = & 
\displaystyle 
\frac{\cpsir}{\cphi} \frac{Q-n}{\varrho} - \frac{m_1}{\varrho},
& \quad &
\displaystyle
\ft_2(\varrho) & = & 
\displaystyle
-\frac{\cpsir}{\cphi}\frac{Q-n}{\varrho} +\frac{m_2}{\varrho}, \\
\vsep
\displaystyle
\ft_3(\varrho) & = & 
\displaystyle
\frac{\cpsil}{\cphi}\frac{Q-n}{\varrho} - \frac{m_3}{\varrho},
& \quad &
\displaystyle
\ft_4(\varrho) & = &
\displaystyle
-\frac{\cpsil}{\cphi}\frac{Q-n}{\varrho} + \frac{m_4}{\varrho}.
\end{array}
\end{equation}
The spinor field ${\Chi}$ verifies the same equations apart from the
fact that, due to its coupling to $\Phi^\dag$ [see
Eq.~(\ref{chilagrangian})], it is necessary to transform $n
\rightarrow -n$.

As was originally found by Jackiw and Rossi~\cite{jackiwrossi} and
Witten~\cite{witten}, there are always $n$ normalizable zero energy
solutions of the Dirac operator in the vortex which allow fermions to
propagate at the speed of light in the ``$-z$'' and ``$+z$,'' say,
directions, for the $\Psi$ and $\Chi$ fields, respectively. These
solutions are found to be eigenvectors of the $\gamma^0 \gamma^3$
operator and are clearly obtained from the above equations by setting
the consistency angular relationships $m_1-1=m_4+n$ and $m_2+1=m_3+n$,
those leading to the zero mode dispersion relation $\miv=0
\Leftrightarrow \omega=\pm k$. Note that only one eigenstate of
$\gamma^0 \gamma^3$ end up being normalizable for each kind of chiral
coupling to the Higgs field, and thus the relevant dispersion
relations reduce to $\omega=-k$ and $\omega=k$, for the $\Psi$ and
$\Chi$ zero modes, respectively~\cite{ringeval}.

Such zero modes have a simple interpretation: since the Higgs field
vanishes in the string core, the mass term $M_\Ferm$ in
Eq.~(\ref{fermmvt}) vanishes too, and the fermions trapped in have
zero mass. As a result, they propagate at the speed of light and they
verify the dispersion relations $\omega= k$ or $\omega=-k$.

\subsection{Massive trapped waves}

However, it is also possible \emph{a priori}, for the trapped
fermions, to explore outer regions surrounding the string core where
the Higgs field takes nonexactly vanishing values. In practice, this
is achieved by means of a nonvanishing fermion angular momentum, which
will lead to a nonvanishing effective mass $\miv^2=\omega^2 -k^2 \neq
0$. For the $\Psi$ field, such massive solutions of the equations of
motion (\ref{systadim}) can only be obtained for four-dimensional
solutions, in order to ease the zero mode constraint $\omega=\pm
k$. The required angular consistency relations therefore read
\begin{equation}
\label{angconst}
m=m_1=m_2+1=m_3+n=m_4+n+1.
\end{equation}
Similarly, the angular dependence of $\Chi$ field has to verify
analogous conditions with the transformation $n \rightarrow -n$. It
was previously shown numerically that the Abelian Higgs model with one
Weyl spinor always admits such kind of normalizable solutions
\cite{davisC}. In the following, massive solutions for Dirac spinors
are numerically derived for our model and shown to exist for a wide
range of fermion charges and coupling constants.

\subsubsection{Analytical considerations}

Some interesting analytical asymptotic behaviors of these modes have
been previously studied~\cite{ringeval,davisC}. In particular, there
are only two degenerate \emph{normalizable} eigensolutions of
Eqs.~(\ref{systadim}) at infinity. Since the Higgs field goes to its
constant vacuum expectation value and the gauge coupling functions
vanish, we found the eigensolutions to scale as $\exp{(\pm \Omega
\varrho)}$, with
\begin{equation}
\Omega = \sqrt{\frac{m_{\mathrm{f}}^2-\miv^2}{m_{\mathrm{h}}^2}}.
\end{equation}
First, note that in order to have decreasing solutions at infinity,
the mass of the trapped modes $\miv$ has to be less than the fermion
vacuum mass $m_{\mathrm{f}}$, as intuitively expected (for
$\miv>m_{\mathrm{f}}$, one recovers the oscillating behaviour that is
typical of free particle solutions). Moreover, from Cauchy theorem,
two degrees of freedom can be set in order to keep only the two well
defined solution at infinity. On the other hand, by looking at the
power-law expansion of both system and solutions near the string
core~\cite{ringeval,jackiwrossi}, only two such solutions are also
found to be normalizable. More precisely, normalizability of each
eigensolution at $\varrho=0$ leads to one condition on the winding
numbers $m_i$ of each spinorial component $\xi_i$. Moreover, in order
for the fermion field to be well defined by rotation around the
string, each spinorial component $\xi_i$ with nonzero winding number
$m_i$ has to vanish in the string core, and so behaves like a
positive power of the radial distance to the core. The analytical
expression of the eigensolutions near $\varrho=0$
reads~\cite{ringeval}
\begin{equation}
\label{corepairs}
\begin{array}{ccccccc}
\left(
\begin{array}{l}
\xi_1 \\
\xi_2 \\
\xi_3 \\
\xi_4
\end{array}
\right)_{s_1}
 & \sim &
\left(
\begin{array}{l}
a_1 \varrho^{-m} \\
a_2(a_1) \varrho^{-m+1} \\
a_3(a_1) \varrho^{-m+|n|+2} \\
a_4(a_1) \varrho^{-m+|n|+1}
\end{array}
\right),
& \quad &
\left(
\begin{array}{l}
\xi_1 \\
\xi_2 \\
\xi_3 \\
\xi_4
\end{array}
\right)_{s_2}
& \sim &
\left(
\begin{array}{l}
a_1 \varrho^{m+|n|-n} \\
a_2(a_1) \varrho^{m+|n|-n+1} \\
a_3(a_1) \varrho^{m-n} \\
a_4(a_1) \varrho^{m-n-1}
\end{array}
\right),
\\ \\
\left(
\begin{array}{l}
\xi_1 \\
\xi_2 \\
\xi_3 \\
\xi_4
\end{array}
\right)_{s_3}
 & \sim &
\left(
\begin{array}{l}
a_1 \varrho^{m} \\
a_2(a_1) \varrho^{m-1} \\
a_3(a_1) \varrho^{m+|n|} \\
a_4(a_1) \varrho^{m+|n|+1}
\end{array}
\right),
& \quad &
\left(
\begin{array}{l}
\xi_1 \\
\xi_2 \\
\xi_3 \\
\xi_4
\end{array}
\right)_{s_4}
& \sim &
\left(
\begin{array}{l}
a_1 \varrho^{-m+|n|+n+2} \\
a_2(a_1) \varrho^{-m+|n|+n+1} \\
a_3(a_1) \varrho^{-m+n} \\
a_4(a_1) \varrho^{-m+n+1}
\end{array}
\right).
\end{array}
\end{equation}
The normalizability condition for the four eigensolutions can be
summarized by 
\begin{equation}
\sup{(0,n)} < m < \inf{(1,1+n)},
\end{equation}
and so, for any value of $m$ there are only two conditions
satisfied. However, from the consistency angular conditions on each
spinorial components, only three pairs of solutions are acceptable
near the string. Assuming $n>0$, if $m \le 0$ then only the pair
$(s_1,s_4)$ is both normalizable and well defined by rotation around
the vortex, similarly for $m \ge n+1$ the relevant solutions are
$(s_2,s_3)$, whereas for $1 \le m \le n$, they are $(s_3,s_4)$.  As a
result, the two remaining degrees of freedom can be set to get only
these pairs near the string core for a given value of $m$, but there
is no reason that they should match with the two normalizable
solutions at infinity. In order to realize this matching we have to
fine tune another parameter which turns out to be the mass of the
modes, $\miv$. As expected for bound states, this mass is therefore
necessarily quantized. Note at this point that $\miv=0$ is, in such a
procedure, nothing but a particular case of the general solution here
presented. The three different pairs of well defined solutions at the
origin suggest that there are three kinds of similar massive bound
states in the vortex, according to the values of the winding number
$m$. Intuitively, the more the field winds around the string, the
farther the particle explores regions surrounding the core due to the
higher values taken by its angular momentum, meaning the largest the
extension of its wave function is, the more it acquires mass from
coupling to a nonexactly vanishing Higgs field. As a result, the
lowest massive modes will certainly be obtained from values of $m$
which correspond to vanishing winding numbers $m_i$.

\subsubsection{Symmetries}
\label{symmetries}

In the following, the equations of motion (\ref{systadim}) will be
summarized in the form $\syse_i^j\at_j=0$, with implicit summation
implied over repeated indices.

The first symmetry is obtained from the complex conjugation of the
equations of motion (\ref{systadim}). Since complex conjugation does
not modify Eqs.~(\ref{systadim}), once the angular consistency
relations (\ref{angconst}) are set, there is an arbitrary complex
phase in the choice of solutions, and it will be sufficient to look
for real rescaled spinorial components $\at_i$.

There is another symmetry between the positive and negative energy
solutions of the equations of motion (\ref{systadim}) that may be
useful. With the label $\varepsilon=\pm $ for particle and
antiparticle states, respectively, one has
\begin{eqnarray}
\sysp_i^j \at_{j_+}=0 &\quad \Rightarrow \quad & \sysm_i^j\at_{j_-}=0,
\end{eqnarray}
provided
\begin{eqnarray}
\at_{i_-} & = & \left(\gamma^0 \gamma^3\right)_i^j \at_{j_+}.
\end{eqnarray}
As a result, the negative energy solutions are obtained from the
positive ones by the action of the $\gamma^0 \gamma^3$ operator, thereby
generalizing the properties of the zero modes which were precisely
found as eigenstates of this
operator~\cite{witten,ringeval,jackiwrossi}.

The last symmetry concerns the gauge coupling functions $\ft_i$. Under
the transformations
\begin{equation}
\label{symcphiwind}
\begin{array}{lllll}
\displaystyle
m & \rightarrow &\widehat{m} & = & n+1-m, \\
\displaystyle
\cpsil & \rightarrow & \widehat{c}_{\psi_{\mathrm{L}}} & = & -\cpsir, \\
\displaystyle
\cpsir & \rightarrow & \widehat{c}_{\psi_{\mathrm{R}}} & = & -\cpsil,
\end{array}
\end{equation}
the gauge functions $\ft_i$, in Eqs.~(\ref{systadim}), are simply
swapped according to $ \ft_1 \leftrightarrow \ft_4$ and $\ft_2
\leftrightarrow \ft_3$. As a result, for every $\at$ solution found
at given $\cpsil$ and $m$, there is another solution $\ah$, with
charge $\widehat{c}_{\psi_{\mathrm{L}}}=\cphi-\cpsil$ and winding number
$\widehat{m}=n+1-m$, namely
\begin{equation}
\label{symcphi}
\begin{array}{ccccccc}
\vsep
\displaystyle
\ah_1(\varrho) & = &  \at_4(\varrho),
& \quad &
\displaystyle
\ah_2(\varrho) & = & \at_3(\varrho), \\
\vsep
\displaystyle
\ah_3(\varrho) & = &  \at_2(\varrho),
& \quad &
\displaystyle
\ah_4(\varrho) & = & \at_1(\varrho).
\end{array}
\end{equation}
Note that the particular case
$\cpsil=\widehat{c}_{\psi_{\mathrm{L}}}=\cphi/2$ appears as a frontier
separating two symmetric kinds of solutions with two differents
winding numbers lying on both sides of $m=(n+1)/2$. As a result, the
three different behaviors found above from normalization and angular
consistency conditions seem to reduce to only two, since the domains
where $m \le 0$ and $m \ge n+1$ are actually connected by charge
symmetry in relation to $\cphi/2$.

On the other hand, due to its coupling to the antivortex instead of
the vortex, the equations of motion of the $\Chi$ field are simply
obtained from Eqs.~(\ref{systadim}) by the transformations $\at_i
\rightarrow \bt_j$, $c_{\psi_{\mathrm{L}(\mathrm{R})}} \rightarrow
c_{\chi_{\mathrm{L}(\mathrm{R})}}$, and $m_i\rightarrow \l_i$. The
$l_i$ are the winding numbers of the scaled $\Chi$ spinorial
components, namely the $\bt_i$, and they verify the angular
consistency relations (\ref{angconst}) with $n$ replaced by $-n$ as
previously discussed.  Let us introduce one more transformation on the
$\Psi$ parameters,
\begin{equation}
\begin{array}{lllll}
\displaystyle
m & \rightarrow & \widehat{m} & = & l+n  , \\
\displaystyle
\cpsil & \rightarrow & \widehat{c}_{\psi_{\mathrm{L}}} & = & \cchir, \\
\displaystyle
\cpsir & \rightarrow & \widehat{c}_{\psi_{\mathrm{R}}} & = & \cchil.
\end{array}
\end{equation}
Naming $\gt_i$ the scaled gauge coupling functions of the $\Chi$ spinor,
the $\Psi$ ones are found to transform according to $\ft_1 \rightarrow
\gt_3$, $\ft_2 \rightarrow \gt_4$, $\ft_3 \rightarrow \gt_1$, and
$\ft_4 \rightarrow \gt_2$. Thus, if the $\at$ are solutions of the
$\Psi$ equations of motion (\ref{systadim}), with $m$ winding number
and $\cpsil$ charge, then there exist $\bt$ solutions for the $\Chi$
field with same mass $\miv$, provided $l=m-n$ and
$\cchil=\cpsir=\cpsil-\cphi$, and they read
\begin{equation}
\label{sympsichi}
\begin{array}{ccccccc}
\vsep
\displaystyle
\bt_1(\varrho) & = & \at_3(\varrho),
& \quad &
\displaystyle
\bt_2(\varrho) & = & -\at_4(\varrho), \\
\vsep
\displaystyle
\bt_3(\varrho) & = & \at_1(\varrho),
& \quad &
\displaystyle
\bt_4(\varrho) & = & -\at_2(\varrho).
\end{array}
\end{equation}
Owing to these symmetries, it is sufficient to study the $\Psi$
equations of motion (\ref{systadim}), for various values of the
winding number $m$ and for left-handed part charges, namely $\cpsil$,
higher or equal than $\cphi/2$.

\subsubsection{Numerical methods}

\begin{figure}
\begin{center}
\epsfig{file=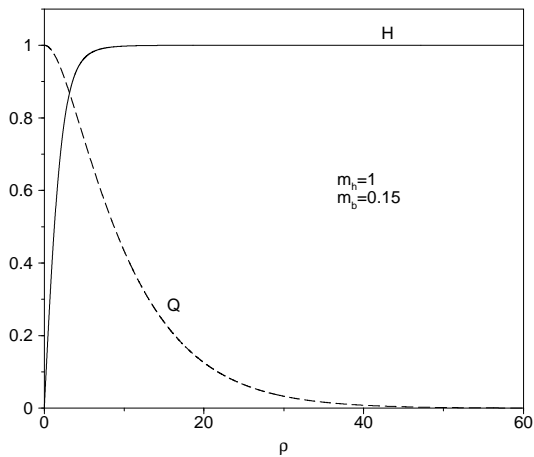,width=7cm}
\caption{The solutions of the field equations for the vortex background. The
Higgs field, $H$, takes its vacuum expectation value at infinity and the
gauge bosons condensate in the vortex.}
\label{figback}
\end{center}
\end{figure}

In order to compute the relevant massive wave solutions for the $\Psi$
fermions on the string, it is necessary to solve first the vortex
background. At zeroth order, neglecting the back reaction of the
fermionic currents, and in terms of the dimensionless fields and
parameters, the equations of motion for the string forming Higgs and
gauge fields read, from Eq.~(\ref{lagrangien}),
\begin{eqnarray}
\label{tildehiggs}
\frac{\ud^2 H}{\ud \varrho^2}+\frac{1}{\varrho} \frac{\ud H}{\ud
\varrho} & = & \frac{H Q^2}{\varrho^2}+\frac{1}{2}H(H^2-1), \\
\label{tildegauge}
\frac{\ud^2 Q}{\ud \varrho^2} -\frac{1}{\varrho}\ \frac{\ud Q}{\ud
\varrho} & = & \frac{m_{\mathrm{b}}^2}{m_{\mathrm{h}}^2}H^2 Q,
\end{eqnarray}
where $m_{\mathrm{b}}=qc_\phi \eta$ is the classical mass of the gauge boson.
The solution of these equations is well known~\cite{neutral,bps,adler}
and shown in Fig.~\ref{figback} for a specific (assumed generic) set
of parameters.

The system of Eqs.~(\ref{systadim}) being linear and involving only
first order derivatives of the spinor components, a Runge-Kutta
numerical method of integration has been used. However, as noted
above, since we are interested only in normalizable solutions, it is
more convenient to perform the resolution from an arbitrary cutoff at
infinity, toward the string core. Let us introduce $\varrho_\infty$,
the cutoff value on the dimensionless radial distance. From the
asymptotic form of Eqs.~(\ref{systadim}) at infinity, and in order to
suppress the exponential growth, the spinorial components $\at_i$ have
to verify
\begin{eqnarray}
\at_1(\varrho_\infty) & = & -\frac{\miv}{\Omega m_{\mathrm{h}}}
\at_2(\varrho_\infty) + \frac{m_{\mathrm{f}}}{\Omega m_{\mathrm{h}}}
\at_4(\varrho_\infty),
\\
\at_3(\varrho_\infty) & = & - \frac{m_{\mathrm{f}}}{\Omega m_{\mathrm{h}}}
\at_2(\varrho_\infty) + \frac{\miv}{\Omega m_{\mathrm{h}}} \at_4(\varrho_\infty).
\end{eqnarray}
These conditions constrain two degrees of freedom, and another one is
fixed by normalization of the wave functions at $\varrho_\infty$. As a
result, only one free parameter can be used yet in order to achieve
the matching between these well defined solutions and the two
normalizable ones near the string core. It will be the case only for
particular values of the mass $\miv$. Numerically, the matching is
performed in two steps. First, by means of the last free parameter,
one of the usually divergent component near the string core is made to
vanish at $\varrho=0$. Obviously, this component is chosen among those
having a nonzero winding number since, in order to be single valued
by rotation around the vortex, it necessarily vanishes at the string
core. Once it is performed, the last divergent component at
$\varrho=0$ is regularized, its turn, by calculating the mass of the
mode $\miv$ leading to a convergent solution. For the range of model
parameters previously defined, the numerical computations thus lead to
the mass of the trapped wave solutions as well as their components as
function of the radial distance to the string core $\at_i(\varrho)$.

\subsubsection{Numerical results}
\label{numresults}

In what follows, the Higgs winding number is assumed fixed to the
value $n=1$, and the range of $\cpsil$ restricted to $\cpsil \ge
\cphi/2$, the other case being derivable from the symmetric properties
discussed above.

\begin{figure}
\begin{center}
\epsfig{file=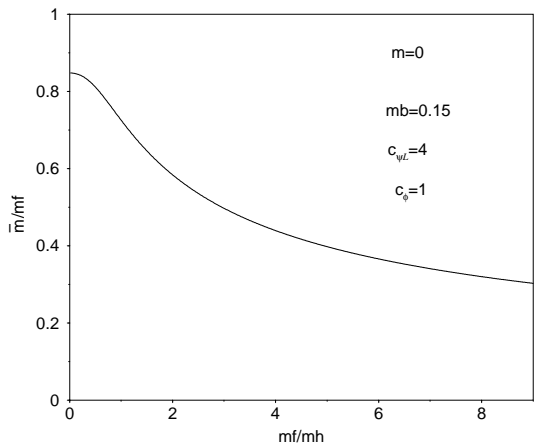,width=7cm}
\caption{The mass of the lowest massive bound state, relative to the
fermion vacuum mass, plotted as function of the coupling constant to
the Higgs field, i.e., the fermion vacuum mass relative to the mass of
the Higgs boson.}
\label{figmfond}
\end{center}
\end{figure}

\begin{figure}
\begin{center}
\epsfig{file=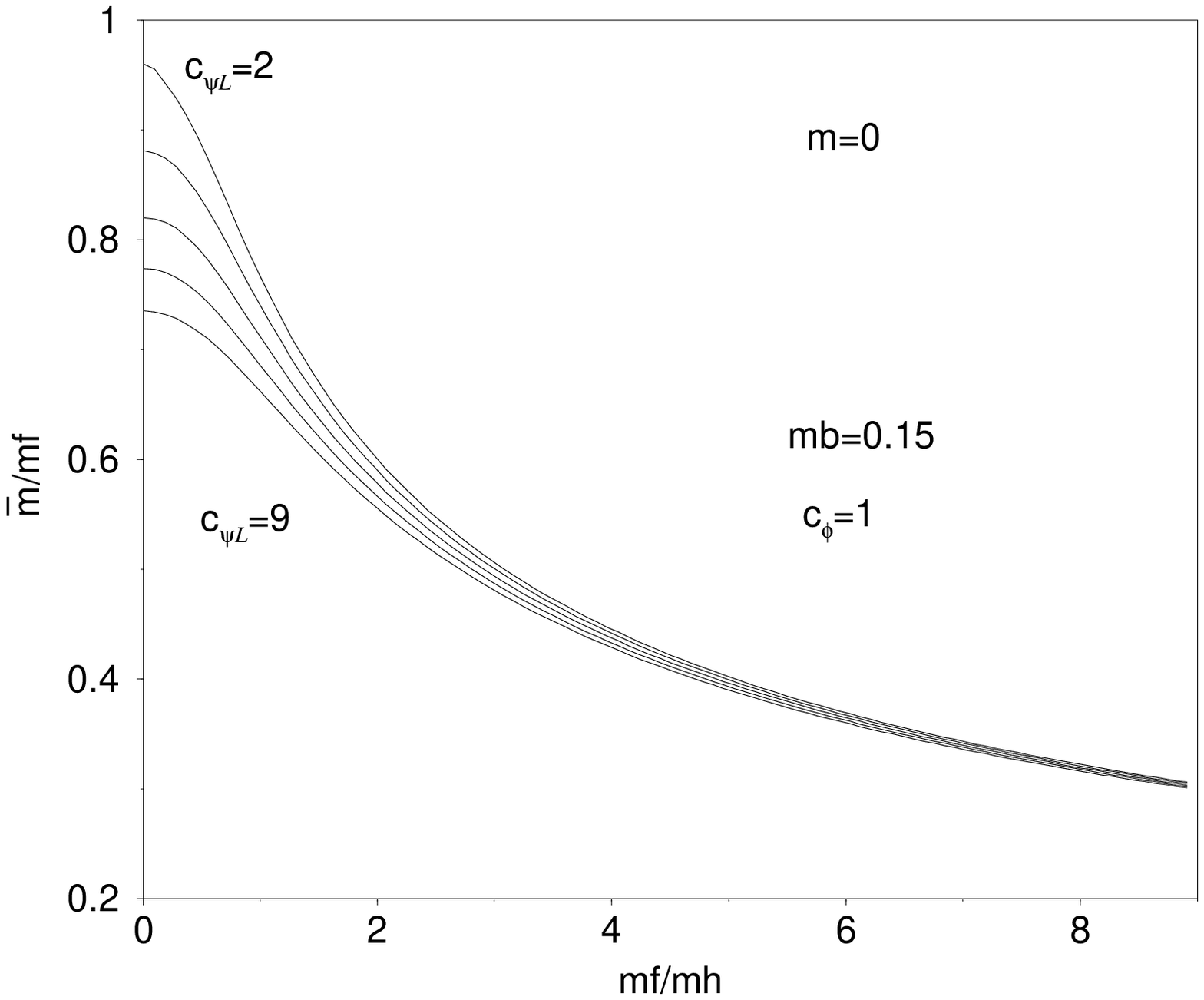,width=7cm}
\caption{The mass of the lowest massive bound state, relative to the
fermion vacuum mass, plotted as function of the coupling constant to
the Higgs field, for several values of the fermionic charges. The
closest $\cpsil$ is to $\cphi/2$, the higher mode mass $\miv$ is. In
the extreme case $\cpsil\sim\cphi/2$, $\miv \sim m_{\mathrm{f}}$ there
is no longer normalizable massive bound state \emph{in the
perturbative sector}.}
\label{figefond}
\end{center}
\end{figure}

\begin{figure}
\begin{center}
\epsfig{file=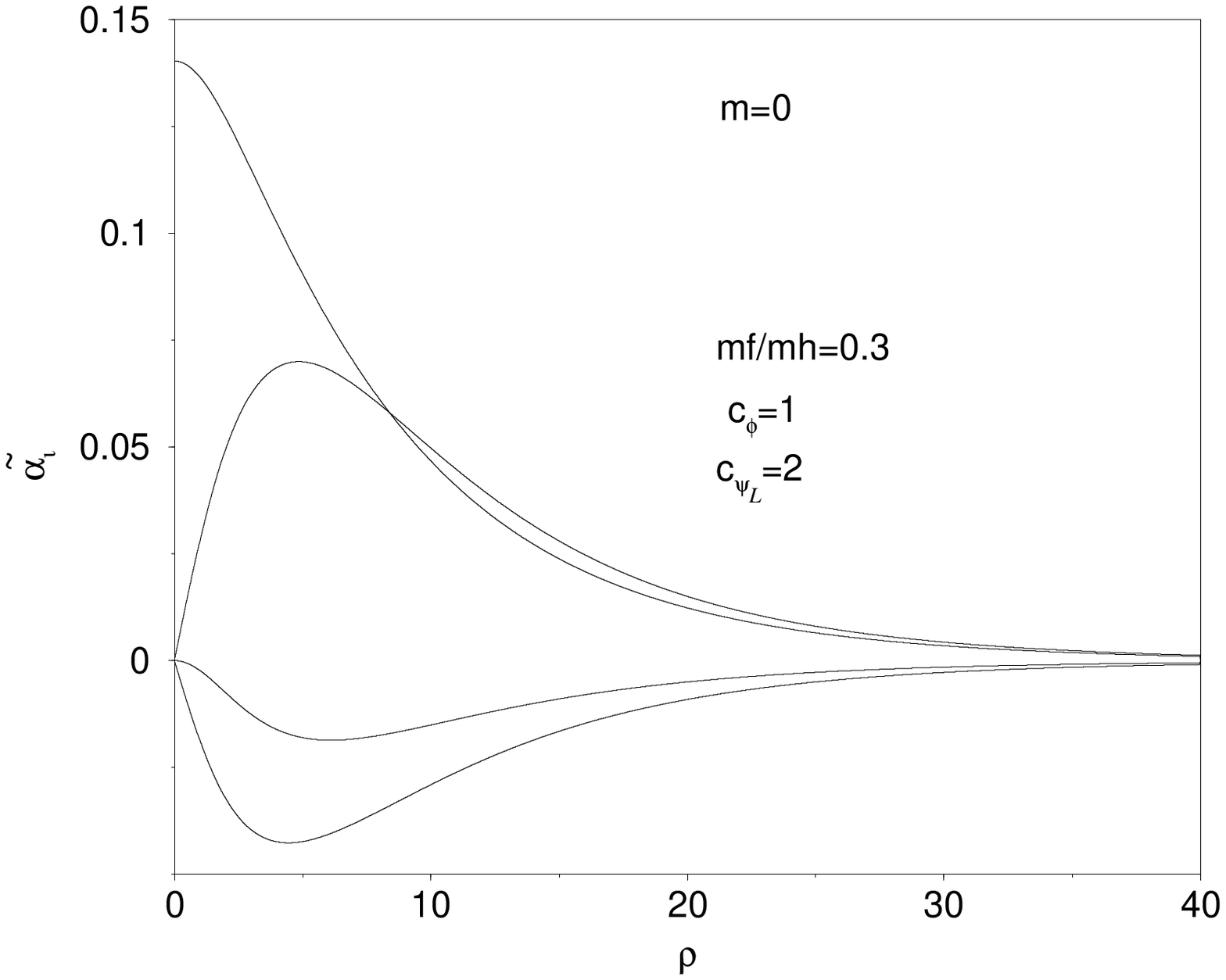,width=7cm}
\epsfig{file=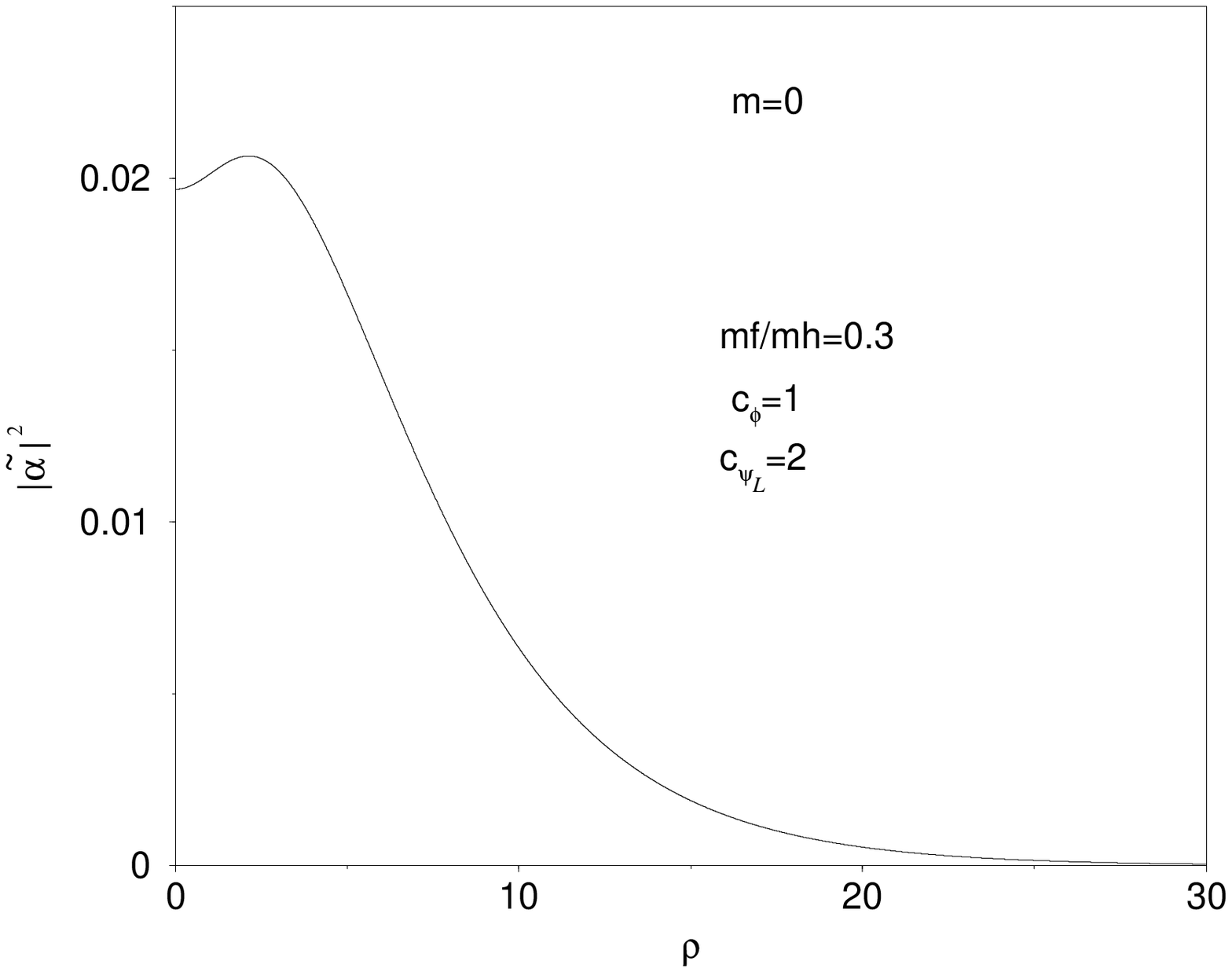,width=7cm}
\caption{The transverse spinorial components of the $\Psi$ field, as
functions of the distance to the string core, for the lowest massive
bound state. The transverse normalized probability density is also
plotted and takes its maximum value nearby the string core, as
expected for massive bound states exploring the neighborhood of the
core by means of nonvanishing angular momentum. Note that for
$m=0$, one spinorial component behaves like a zero mode one, i.e.,
it condenses in the string core contrary to the others.}
\label{figprobafond}
\end{center}
\end{figure}

The first results concern the ``perturbative sector'' where the
fermion vacuum mass verifies $m_{\mathrm{f}}<m_{\mathrm{h}}$, or
equivalently, for a smaller Yukawa coupling constant than the Higgs
self-coupling, i.e., $g<\sqrt{\lambda}$. In this case, for reasonable
values of the fermion charges, i.e., of the same order of magnitude
than the Higgs one $\cpsil \gtrsim \cphi/2$, only one normalizable
massive bound state is found with null winding number $m=0$. As a
result, by means of transformations~(\ref{symcphiwind}), there are
also symmetric modes for $\cpsil \le \cphi/2$, with winding number
$m=2$. The dependency of the mode mass $\miv$ with the fermion vacuum
mass and charges (i.e., the coupling constants to Higgs and gauge
fields) is plotted in Fig.~\ref{figmfond} and Fig.~\ref{figefond}. The
study has been also extended to the nonperturbative sector where this
massive mode thus appears as the lowest massive bound state. First, it
is found that the mass of the trapped mode always decreases with
respect to the coupling constant, i.e, with the fermion vacuum mass
$m_{\mathrm{f}}$. Moreover, for small values of
$m_{\mathrm{f}}/m_{\mathrm{h}}$, the derivative of the curve
$\miv(m_{\mathrm{f}}/m_{\mathrm{h}})$ vanishes near the origin (see
Fig.~\ref{figmfond}). As a result, the mass modes in the full
perturbative sector does not depend on the coupling constant to the
Higgs field, at first order. On the other hand, Fig.~\ref{figefond}
shows that the mass of the bound state hardy depends at all on its
coupling with the gauge field (i.e., on the charges $\cpsil$) in the
nonperturbative sector, where all the curves have the same asymptotic
behavior. Near the origin, the closest $\cpsil$ is to $\cphi/2$, the
higher mode mass $\miv$ is. In the particular limiting case $\cpsil
\sim \cphi/2$, there is no normalizable massive bound state, and as
can be seen in Fig.~\ref{figefond}, already for $\cpsil/\cphi=2$, the
mode mass is close to $m_{\mathrm{f}}$. It is not surprising since, as
it was above noted, $\cpsil=\cphi/2$ is a frontier between two kinds
of solutions with different winding numbers, and thus, at this point,
the ``normalizable'' winding numbers are not well defined. Note that
this is only true if $m\neq (n+1)/2$ as it is the case here in the
perturbative sector with $n=1$ and $m=0$.
The normalized scaled spinorial components $\at(\varrho)$ have been
plotted in Fig.~\ref{figprobafond} for the lowest massive bound
state, with the normalization
\begin{equation}
\int{\varrho \,\ud \varrho\, \at_i^{\dag} \at_i} = 1.
\end{equation}
The corresponding transverse probability density has also been plotted
in Fig.~\ref{figprobafond}. Note that the massive mode wave function
is larger around the string rather on it, as expected for a nonvanishing
angular momentum.

\begin{figure}
\begin{center}
\epsfig{file=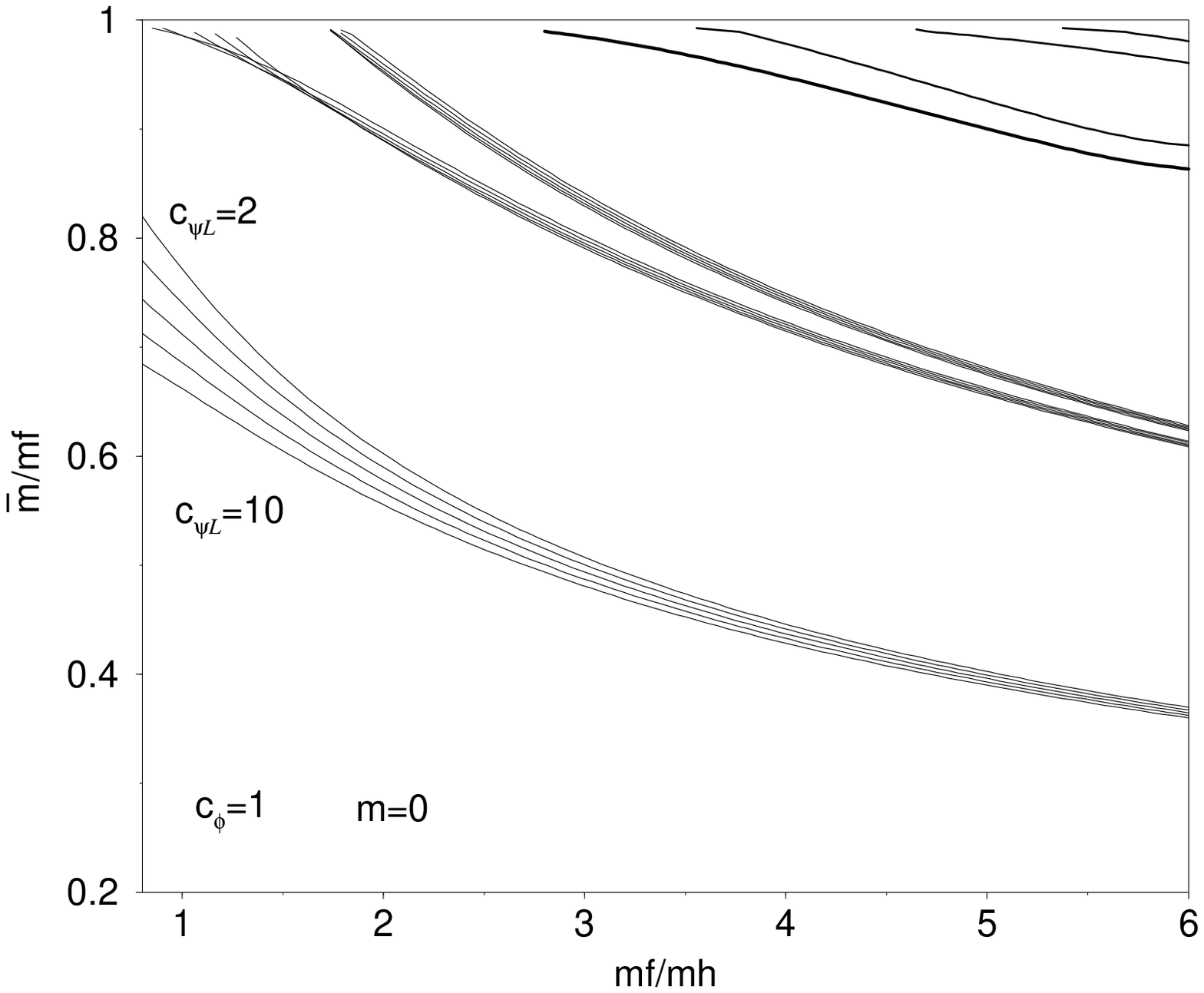,width=7cm}
\caption{The evolution of the mass spectrum for $m=0$ winding number,
as function of the coupling constants. Each main branch represents one
massive mode whereas the substructures show its evolution with respect
to the fermion charge $\cpsil$. Five values of the fermion charge have
been plotted, from $\cpsil=2$ to $\cpsil=10$, and the spectrum has
been computed only in the nonperturbative sector, since only the
lowest mode exists for lower values of $m_{\mathrm{f}}/m_{\mathrm{h}}$
(see Fig.~\ref{figefond}). As expected, all the modes have mass $\miv$
decreasing with their coupling constant to the Higgs field. Moreover,
the substructures show that, for sufficiently large values of
$m_{\mathrm{f}}/m_{\mathrm{h}}$, the mode mass is a decreasing
function of the charge $\cpsil$. However, note that this behavior can
be inverted for some modes close to their appearance region, as it is
the case for the second one.}
\label{figmasspect}
\end{center}
\end{figure}

\begin{figure}
\begin{center}
\epsfig{file=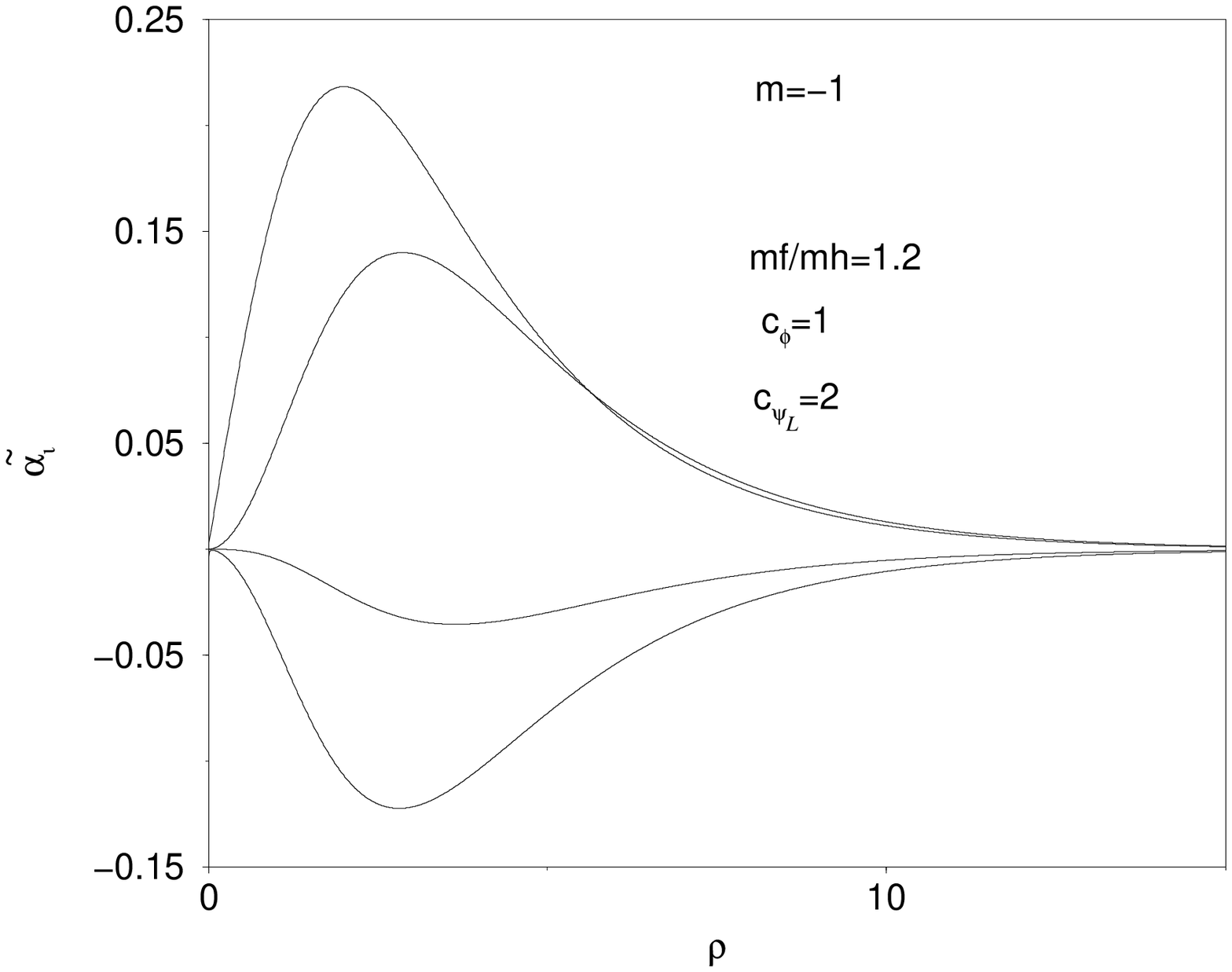,width=7cm}
\epsfig{file=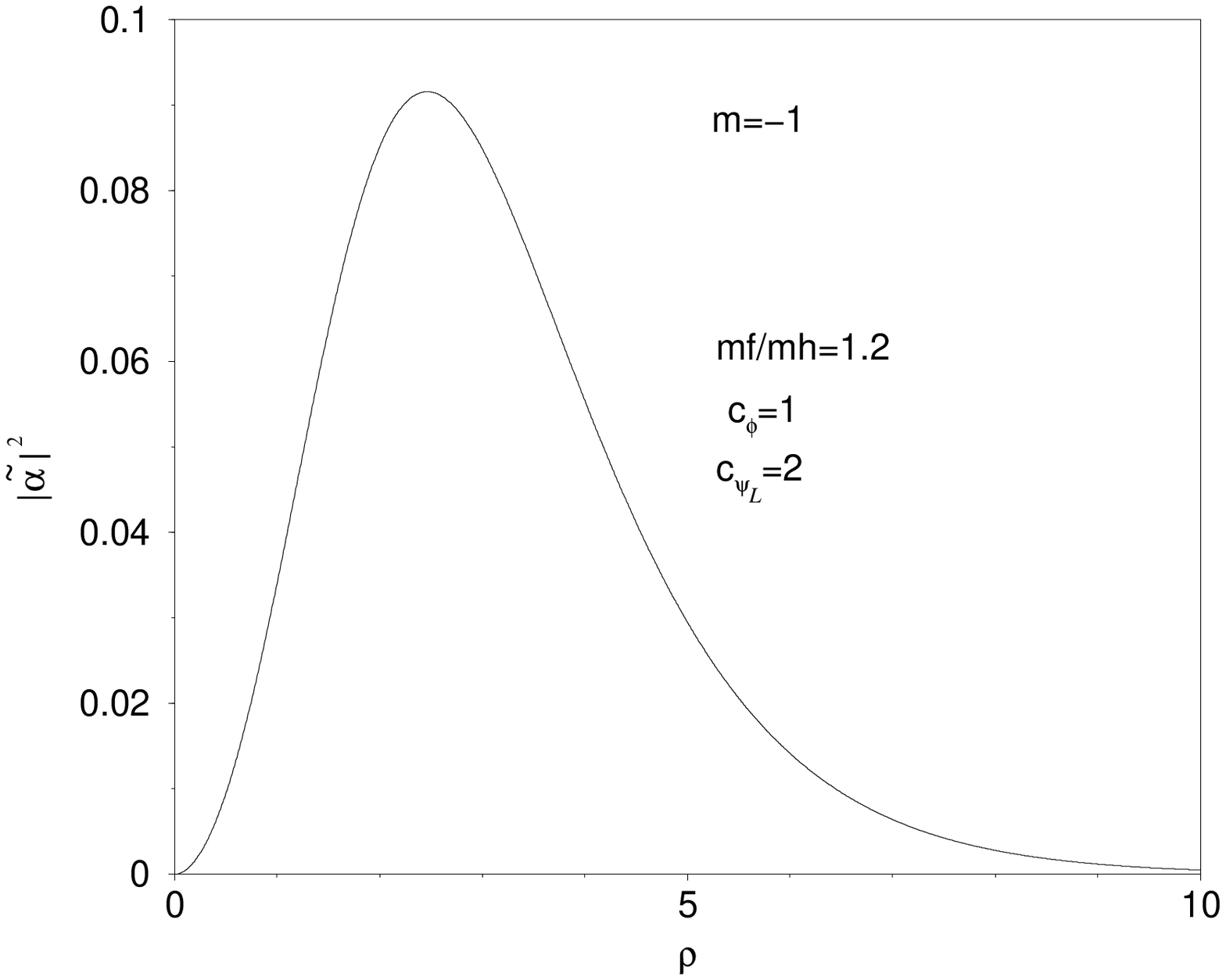,width=7cm}
\caption{The transverse spinorial components of the $\Psi$ field,
plotted as functions of the distance to the string core, for the
$m=-1$ lowest massive bound state. The transverse normalized
probability density is also plotted and vanishes in the string
core. In this case, all components of the spinor wind around the
string and the corresponding mode is thus centrifugally confined in a
shell nearby the core, as expected for a nonzero angular momentum
eigenstate.}
\label{figprobafond1}
\end{center}
\end{figure}

\begin{figure}
\begin{center}
\epsfig{file=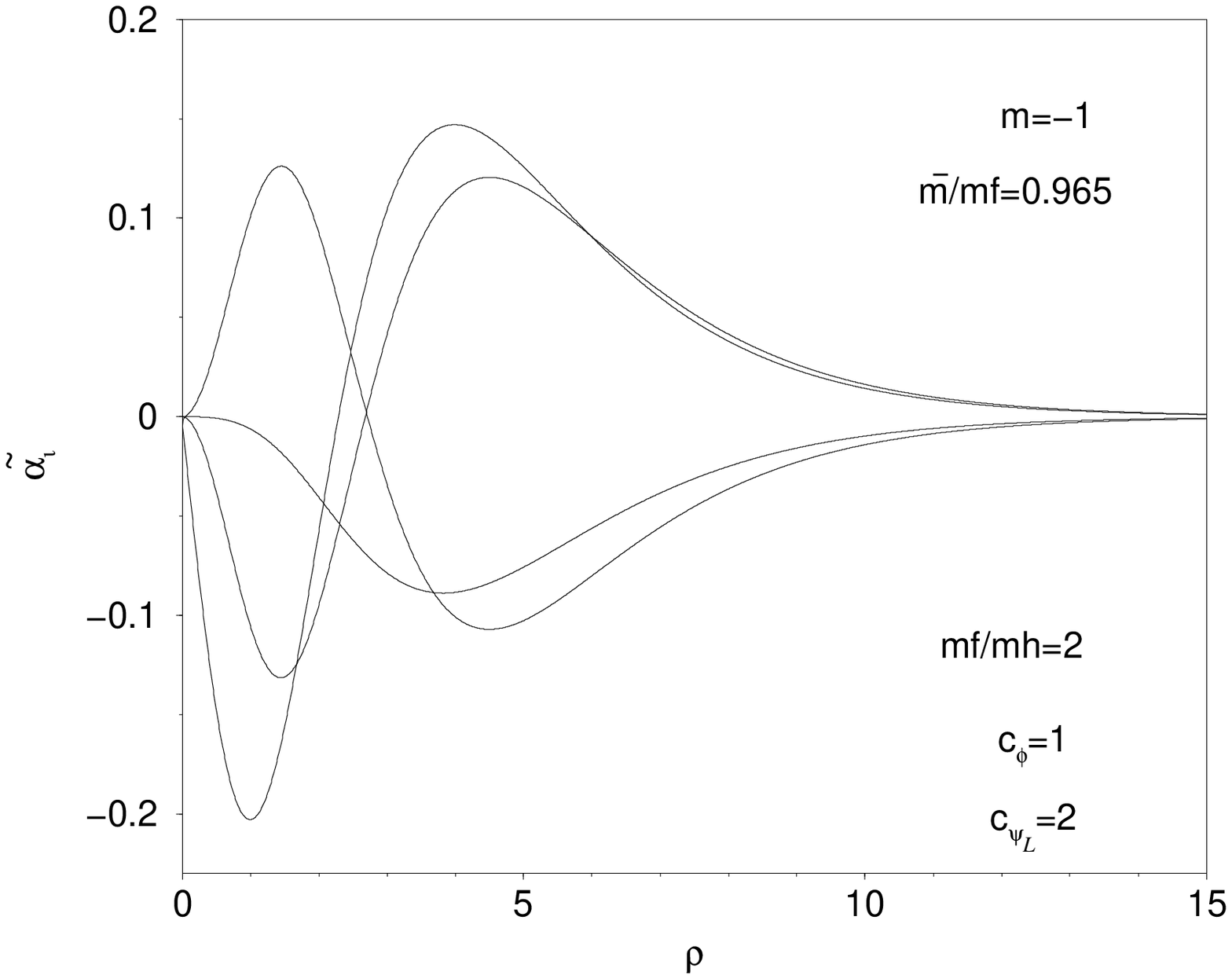,width=7cm}
\epsfig{file=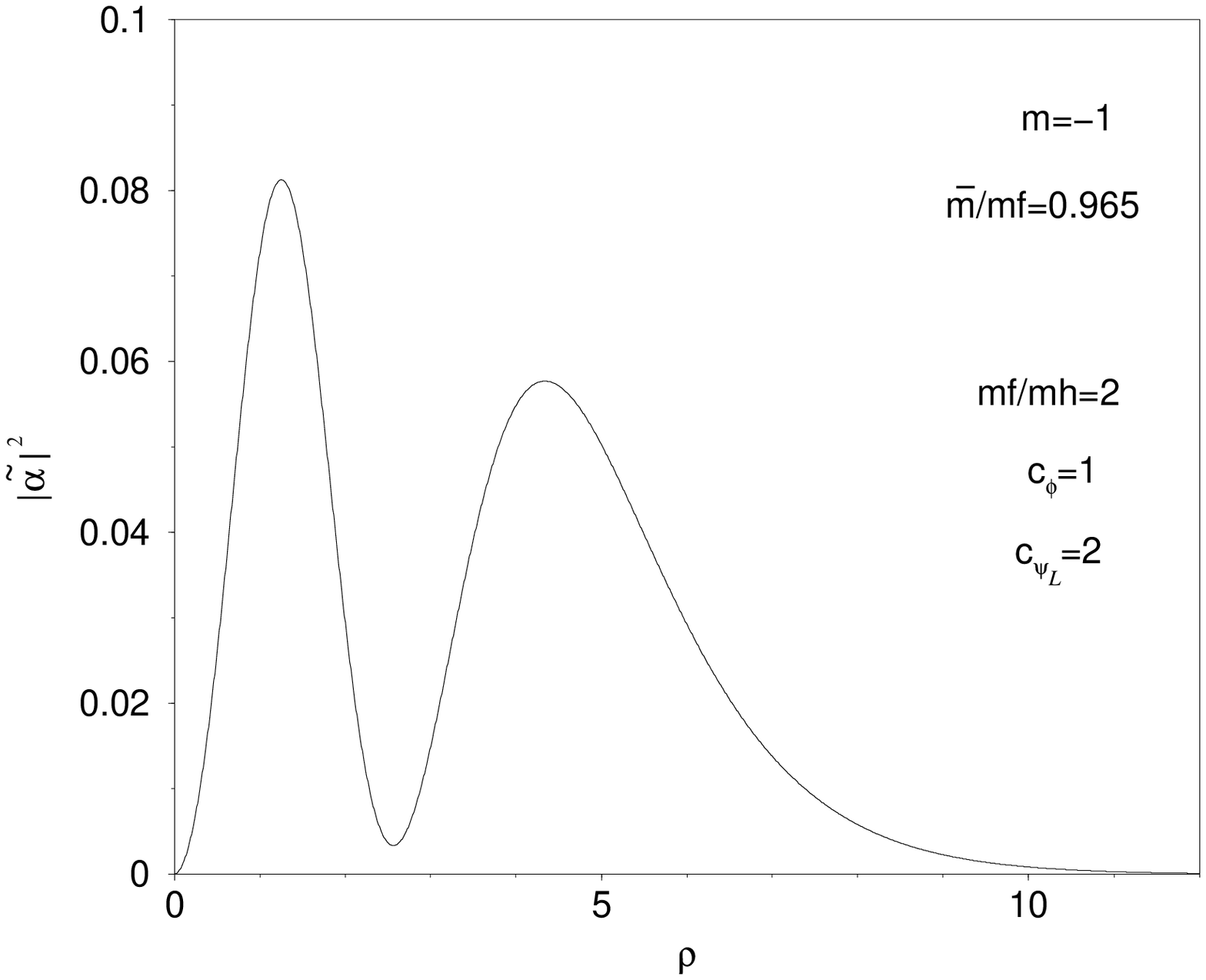,width=7cm}\\
\epsfig{file=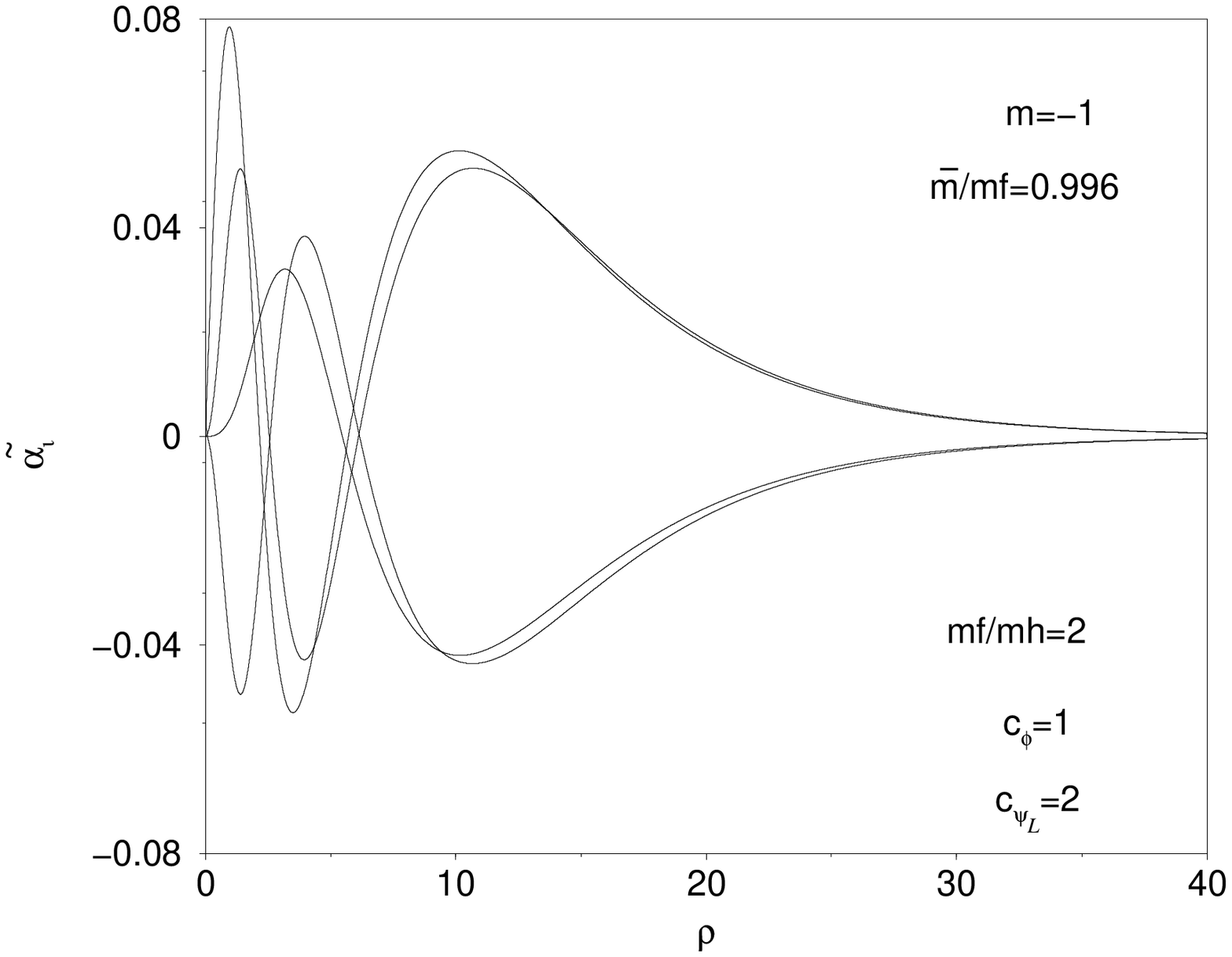,width=7cm}
\epsfig{file=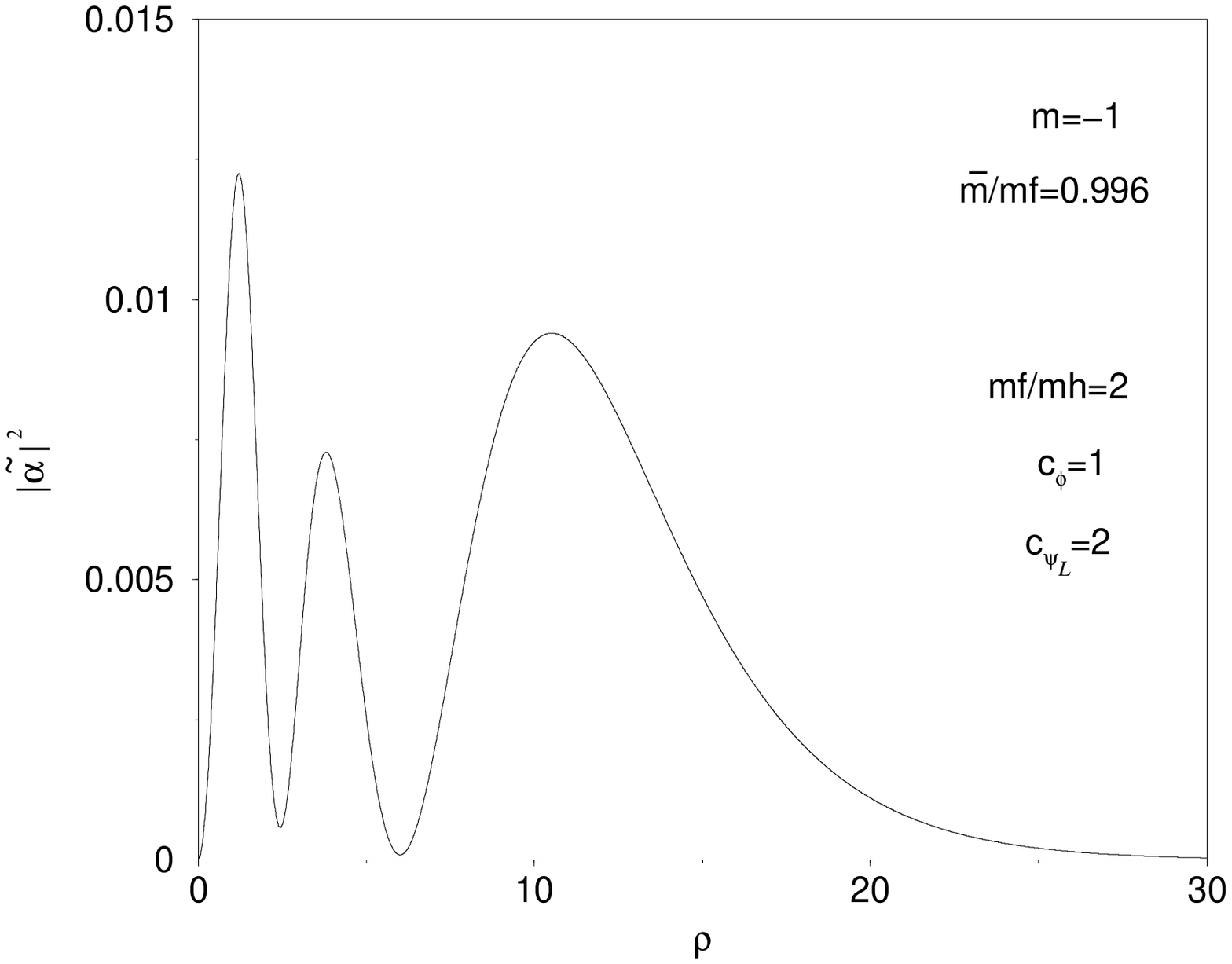,width=7cm}
\caption{The components of the $\Psi$ field as function of the radial
distance, and the corresponding transverse probability densities. The
curves have been computed for $m=-1$ winding number, and for two
additional incoming modes in the nonperturbative sector. As expected,
interferences fringes appear from the nonzero angular momentum
eigenvalues of these modes.}
\label{figprobaex}
\end{center}
\end{figure}

\begin{figure}
\begin{center}
\epsfig{file=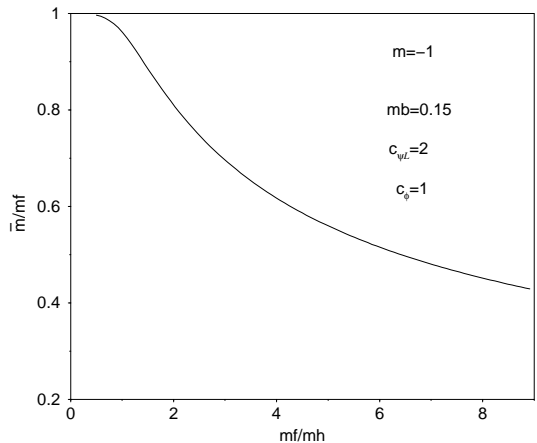,width=7cm}
\caption{The mass of the $m=-1$ lowest massive bound state, relative
to the fermion vacuum mass, with respect to the coupling constant to
the Higgs field. Note that the mode does not exist in the perturbative
sector.}
\label{figmassfond1}
\end{center}
\end{figure}

The nonperturbative cases with $m_{\mathrm{f}}>m_{\mathrm{h}}$,
involve much more massive bound states. First, another mode appears in
addition to the previous one, with the same winding number. Because of
the fact that $\miv$ decreases with $m_{\mathrm{f}}$ [see
Fig.~\ref{figmfond}], for higher values of $m_{\mathrm{f}}$, another
mode comes into the normalizable mass range. Since normalizability at
infinity requires $\miv<m_{\mathrm{f}}$, the number of massive modes
increases with the value of $m_{\mathrm{f}}$. Moreover, there are also
solutions involving all the other possible winding numbers. The
evolution of the mass spectrum, for winding number $m=0$, and with
respect to the coupling constant to the Higgs and gauge fields is
plotted in Fig.~\ref{figmasspect}. The behavior of each mass is the
same as that of the lowest mode previously studied, the new properties
resulting only in the appearance of new states for higher values of
the fermion vacuum mass $m_{\mathrm{f}}$, as found for two-dimensional
Weyl spinors in Ref.~\cite{davisC}.
Physically, the additional massive modes at a given winding number can
be interpreted as normalizable eigenstates of the angular momentum
operator in the vortex potential, with higher eigenvalues. From
Fig.~\ref{figprobafond} and Fig.~\ref{figprobafond1}, one can see
that for each value of $m$, the lowest massive state is confined
around the string with a transverse probability density showing only
one peak whereas the higher massive modes have transverse probability
density profiles with an increasing number of maxima, as can
be seen in Fig.~\ref{figprobaex}. In fact, as for the structure of
atomic spectra, the two spatial degrees of freedom of the attractive
potential certainly lead to two quantum numbers labeling the observable
eigenstates, one of them being clearly $m$, and the other appearing
through the number of zeros of the spinorial components, or, equivalently,
the number of maxima of the associated transverse probability density.
The massive modes with higher winding numbers behave in the same way.
However, they exist only for nonzero values of the coupling constant
$m_{\mathrm{f}}/m_{\mathrm{h}}$, this one increasing with the value of
the winding number $m$. The scaled spinorial components and the
transverse normalized probability density of the lowest massive bound
state with next $m=-1$ winding number are plotted in
Fig.~\ref{figprobafond1}. They are found to be normalizable for
coupling constant $m_{\mathrm{f}}/m_{\mathrm{h}} \gtrsim 0.5$ when
$\cpsil/\cphi=2$, as can be seen in Fig.~\ref{figmassfond1}. Contrary
to the $m=0$ lowest massive state, all spinorial components wind
around the string, and the transverse probability of finding such a
mode vanishes in the string core, as expected for a nonzero angular
momentum eigenstate. Obviously, this is also true for all higher
values of $m$, as for the $m=-2$ massive mode which appears to be
normalizable for $m_{\mathrm{f}}/m_{\mathrm{h}} \gtrsim 1.2$.
It is clear from the numerical results that the fermions can be
trapped in the string in the form of massive bound states, for a wide
range of model parameters. The only exception takes place for fermion
charges close to the particular value $\cpsil=\cphi/2$ where there is
no normalizable massive bound state in the perturbative sector. Note,
again, that all the previous results are also relevant for the
massive modes with symmetric winding numbers $\widehat{m}=n+1-m$, as
well as for the $\Chi$ spinor field and for the antiparticle states of
both $\Psi$ and $\Chi$ fermions.

\section{Fock space for massive modes}
\label{quantization}
The existence of massive trapped waves requires that the quantum
state space~\cite{ringeval} be enlarged to include them. For each
normalizable mode with mass $\miv$, a two-dimensional Fock space can
be constructed by spinor field expansion over the relevant massive
plane waves. The full quantum theory can therefore be obtained from
tensorial product of the different Fock spaces belonging to their own
mass representation, together with the Fock space associated with the
zero modes~\cite{ringeval}. As a first step toward a full theory, we
will only consider the plane waves associated with one massive mode of
mass $\miv$.

\subsection{Quantum field operators}

Quantization is performed through the canonical procedure by
defining creation and annihilation operators satisfying
anticommutation rules. However, the particular structure of the
trapped massive waves yields relationships between longitudinal
quantum operators with nontrivial transverse dependencies.

\subsubsection{Fourier transform}

In the previous section, it was shown that the fermions could
propagate along the string direction with given mass $\miv$ belonging
to the spectrum. From Eq.~(\ref{planeansatz}), setting
\begin{eqnarray}
\Psi_{\mathrm{p}}^{(+)}  = u_\psi(k,r,\theta) \, \ue^{i(\omega t-kz)},
& \quad &
\Psi_{\mathrm{p}}^{(-)}  = v_\psi(k,r,\theta) \, \ue^{-i(\omega t-kz)},
\end{eqnarray}
with
\begin{equation}
\label{dispmass}
\omega=\sqrt{\miv^2+k^2},
\end{equation}
and using the symmetry properties shown in Sec.~(\ref{symmetries}),
the transverse parts of the massive trapped waves for particle
and antiparticle states can be written as
\begin{equation}
\label{psiuv}
\begin{array}{ccccccc}
u_\psi(k,\xper)
 & = &
\left(
\begin{array}{c}
\sqrt{\omega+k}\,\ab_1(r)\, \ue^{-im\theta}\\
i\sqrt{\omega-k}\,\ab_2(r)\, \ue^{-i(m-1)\theta}\\
\sqrt{\omega-k}\,\ab_3(r) \,\ue^{-i(m-n)\theta}\\
i\sqrt{\omega+k}\,\ab_4(r)\,\ue^{-i(m-n-1)\theta}
\end{array}
\right),
& \quad &
v_\psi(k,\xper)
& = &
\left(
\begin{array}{c}
\sqrt{\omega+k}\,\ab_1(r)\, \ue^{-im\theta}\\
-i\sqrt{\omega-k}\,\ab_2(r)\, \ue^{-i(m-1)\theta}\\
-\sqrt{\omega-k}\,\ab_3(r) \,\ue^{-i(m-n)\theta}\\
i\sqrt{\omega+k}\,\ab_4(r) \,\ue^{-i(m-n-1)\theta}
\end{array}
\right),
\end{array}
\end{equation}
with the notations
\begin{eqnarray}
\label{defalphabar}
\xper=(r,\theta), & \quad \textrm{and} \quad &
\ab=\frac{m_{\mathrm{h}}}{\sqrt{2\pi}}\at.
\end{eqnarray}
Contrary to the zero mode case, fermions can now propagate in both
directions of the string, so that the momentum $k$ of the massive
waves can take positive and negative values. As a result, the $\Psi$
field can be Fourier expanded over positive and negative energy states
as
\begin{equation}
\label{psiexpansion}
\Psi=\int{\frac{\ud k}{2\pi 2\omega} \left[b^{\dag}(k)\, u(k,\xper)\,
\ue^{i(\omega t-kz)} + d(k)\, v(k,\xper)\, \ue^{-i(\omega t-kz)} \right]},
\end{equation}
where the subscripts have been omitted. The normalization convention
of the Fourier transform is chosen as in the zero modes
case~\cite{ringeval}, i.e.,
\begin{equation}
\label{delta}
\int{\ud z\, \ue^{i(k-k^\prime)z}} = 2\pi \delta(k-k^\prime).
\end{equation}
Obviously, the $\Chi$ field verifies similar relations with the
transformation $n \rightarrow -n$, as was noted previously.

\subsubsection{Commutation relations}

In order to express the Fourier coefficients $b(k)$ and $d^\dag(k)$ as
function of the spinor field $\Psi$, let us introduce another unit
spinors
\begin{equation}
\label{psiuhvh}
\begin{array}{ccccccc}
\uh(k,\xper)
 & = &
\left(
\begin{array}{c}
\sqrt{\omega+k}\,\ab_3(r)\, \ue^{-im\theta}\\
i\sqrt{\omega-k}\,\ab_4(r)\, \ue^{-i(m-1)\theta}\\
\sqrt{\omega-k}\,\ab_1(r) \,\ue^{-i(m-n)\theta}\\
i\sqrt{\omega+k}\,\ab_2(r)\,\ue^{-i(m-n-1)\theta}
\end{array}
\right),
& \quad &
\vh(k,\xper)
& = &
\left(
\begin{array}{c}
\sqrt{\omega+k}\,\ab_3(r)\, \ue^{-im\theta}\\
-i\sqrt{\omega-k}\,\ab_4(r)\, \ue^{-i(m-1)\theta}\\
-\sqrt{\omega-k}\,\ab_1(r) \,\ue^{-i(m-n)\theta}\\
i\sqrt{\omega+k}\,\ab_2(r) \,\ue^{-i(m-n-1)\theta}
\end{array}
\right).
\end{array}
\end{equation}
They clearly verify $\uh=\gamma^0\gamma^3 \vh$ and from
Eq.~(\ref{psiuv})
\begin{equation}
\label{orthonorm}
\begin{array}{ccccl}
\vsep \displaystyle
\uh^\dag(k)u(k) & = & \vh^\dag(k) v(k) & = &
2\omega \, \nub(r),
\\
\vsep \displaystyle
\uh(k)^\dag v(-k) & = & \vh^\dag(k) u(-k) & = & 0,
\end{array}
\end{equation}
where the dependency with respect to transverse coordinates have been
omitted in order to simplify the notation, and where we introduced the
function
\begin{equation}
\nub(r)=\ab_1(r) \ab_3(r) + \ab_2(r) \ab_4(r).
\end{equation}
From Eqs.~(\ref{psiexpansion}), (\ref{delta}), and (\ref{orthonorm}),
the Fourier coefficients are found to be functions of the $\Psi$ field,
and read
\begin{equation}
\label{fouriercoeff}
\begin{array}{lcl}
\vsep
\displaystyle
b^\dag(k) & = & \displaystyle
\frac{1}{\norm} \int{r\,\ud r\,\ud \theta \,\ud z\,
\ue^{-i(\omega t-kz)} \uh^\dag(k,\xper) \Psi },
\\
\vsep
\displaystyle
d(k) & = & \displaystyle
\frac{1}{\norm} \int{r\,\ud r\,\ud
\theta \,\ud z\, \ue^{i(\omega t-kz)} \vh^\dag(k,\xper) \Psi},
\end{array}
\end{equation}
where we have defined the normalization factor
\begin{equation}
\norm=\int{r\,\ud r\,\ud\theta\, \nub(r)}=\int{\varrho\,\ud\varrho \,
\nut(\varrho)}
\end{equation}
Similarly, the expansion of the $\Psi^\dag$ field on the same positive
and negative energy solutions leads to the definition of its Fourier
coefficients, namely $b(k)$ and $d^{\dag}(k)$. From Eqs.~(\ref{delta})
and (\ref{orthonorm}), they can also be expressed as functions of
$\Psi^\dag$, and verify
\begin{eqnarray}
\label{fourierconj}
b(k)=[b^\dag(k)]^\dag, & \quad \textrm{and} \quad &
d^\dag(k)=[d(k)]^\dag.
\end{eqnarray}

In order to perform a canonical quantization along the string
world sheet, let us postulate the anticommutation rules, \emph{at equal
times}, between the spinor fields
\begin{eqnarray}
\label{psiquantization}
\left\{\Psi_i(t,\vec x), \Psi^{\dag j} (t,\vec x^\prime) \right\} & =
& \delta(z-z^\prime) \left(\Gamma^{0}\right)_i^j(\xper,\xper'),
\end{eqnarray}
where $\Gamma^0$ is a matrix with respect to spinor components whose
utility will be justified later, and which reads
\begin{eqnarray}
\Gamma^0(\xper,\xper') & = & \frac{1}{2 \miv^2} \left(\omega
{\mathrm{I}} - k \gamma^0 \gamma^3\right)\left[u(k,\xper)
u^\dag(k,\xper')+v(k,\xper) v^\dag(k,\xper') \right].
\end{eqnarray}
Note that $\Gamma^0$ does not depend on $\omega$ and $k$. Explicit
calculations show that the first terms involving $\omega$ and $k$ are
mixed with $u(k)$ and $v(k)$, and yield Lorentz invariant
quantities, such as $\miv$. Moreover, $\Gamma^0$ has the following
orthonormalization properties
\begin{equation}
\label{gammaprop}
\begin{array}{lcl}
\vsep
\displaystyle
\uh^\dag(k,\xper)\,\Gamma^0(\xper,\xper')\,\uh(k,\xper')
& = & 2\omega \,\nub(r) \nub(r')
\\
\vsep
\displaystyle
\uh^\dag(k,\xper)\,
\Gamma^0(\xper,\xper')\, \vh(-k,\xper') & = & 0,
\end{array}
\end{equation}
and similar relationships are obtained for $\vh$ by swapping $\uh$ and
$\vh$.

The anticommutation rules for the $\Psi$ Fourier coefficients are
immediately obtained from Eqs.~(\ref{delta}), (\ref{fouriercoeff}),
(\ref{fourierconj}), and Eq.~(\ref{psiquantization}), using the
properties of $\Gamma^0$ in Eq.~(\ref{gammaprop}), and read
\begin{equation}
\label{creatoranticom}
\begin{array}{ccccclll}
\left\{b(k), b^\dag(k^\prime) \right\} & = & \left\{d(k),
d^\dag(k^\prime)
\right\} & = & 2\pi2\omega \delta(k-k^\prime),
\end{array}
\end{equation}
with all the other anticommutators vanishing. As a result, the
Fourier coefficients $b^\dag$ and $d^\dag$ behave as well defined
creation operators, whereas their complex conjugates, $b$ and $d$, act
as annihilation operators of a particle and antiparticle massive
state, respectively.

In order to verify the microcausality of the theory and to justify,
\emph{a posteriori}, the ansatz of Eq.~(\ref{psiquantization}), let us
derive the anticommutator between the quantum field operators $\Psi$
and $\Psi^\dag$, \emph{at any time}. The $\Psi$ expansion in
Eq.~(\ref{psiexpansion}) and its complex conjugate yield
\begin{eqnarray}
\label{fieldsanticom}
\left\{\Psi_i({\bold{x}}), \Psi^{\dag j} ({\bold{x}}^\prime) \right\}
& = &\int{\frac{\ud k\, \ud k^\prime}{(2\pi)^2 2\omega 2\omega'}}
\left[ \left\{b^\dag(k),b(k')\right\} u_i(k,\xper) u^{\dag
j}(k',\xper') \ue^{i[(\omega t -\omega' t')-(kz-k'z')]} + \right.
\nonumber \\ & + & \left.  \left\{d(k),d^\dag(k')\right\} v_i(k,\xper)
v^{\dag j}(k',\xper')\ue^{-i[(\omega t -\omega'
t')-(kz-k'z')]}\right].
\end{eqnarray}
Using Eq.~(\ref{creatoranticom}), this equation simplifies to involve
tensorial products of unit spinors evaluated at the same momentum. It
is therefore convenient to introduce two additional matrices, namely
$\Gamma^3(\xper,\xper')$ and $\M(\xper,\xper')$, which verify
\begin{equation}
\label{tensorsplit}
\begin{array}{lcl}
\vsep \displaystyle u(k,\xper) u^{\dag}(k,\xper') & = & \displaystyle
\omega \Gamma^0(\xper,\xper')-k
\Gamma^3(\xper,\xper')-\M(\xper,\xper') \\ \vsep \displaystyle
v(k,\xper) v^{\dag}(k,\xper') & = & \displaystyle \omega
\Gamma^0(\xper,\xper')-k \Gamma^3(\xper,\xper')+\M(\xper,\xper').
\end{array}
\end{equation}
From Eq.~(\ref{psiuv}), these matrices are simply related to $\Gamma^0$
by
\begin{equation}
\begin{array}{lcl}
\label{gamma03m}
\vsep
\displaystyle
\Gamma^3(\xper,\xper') & = & \Gamma^0(\xper,\xper')\, \gamma^3
\gamma^0,\\
\vsep
\displaystyle
\M(\xper,\xper') & = &
\displaystyle
\Gamma^0(\xper,\xper')\,
\M_{\mathrm{d}}(\xper') \gamma^0,
\end{array}
\end{equation}
where $\M_{\mathrm{d}}(\xper)$ is the diagonal matrix
\begin{eqnarray}
\label{md}
\M_{\mathrm{d}}(\xper) & = & \miv \,{\mathrm{Diag}}
\left(\frac{\ab_3(r)}{\ab_1(r)}
\ue^{-in\theta},\frac{\ab_4(r)}{\ab_2(r)}
\ue^{-in\theta},\frac{\ab_1(r)}{\ab_3(r)}
\ue^{in\theta},\frac{\ab_2(r)}{\ab_4(r)} \ue^{in\theta}\right).
\end{eqnarray}
From Eqs.~(\ref{tensorsplit}) and (\ref{gamma03m}), the anticommutator
(\ref{fieldsanticom}) reduces to
\begin{eqnarray}\label{reducecom}
\left\{\Psi({\bold{x}}), \Psi^{\dag} ({\bold{x}}^\prime) \right\}
& = & \left[\Gamma^0(\xper,\xper')\,i\partial_0 +
\Gamma^3(\xper,\xper')\,
i \partial_3 + \M(\xper,\xper') \right] \,i \Delta(\xpar-\xpar'),
\end{eqnarray}
where $\xpar=(t,z)$, and $\Delta$ is the well-known Pauli Jordan
function which reads
\begin{eqnarray}
i\Delta(\xpar-\xpar') & = & \int{\frac{\ud k}{2\pi 2\omega}
\left[\ue^{-ik(\xpar-\xpar')} - \ue^{ik(\xpar-\xpar')}\right]},
\end{eqnarray}
and vanishes outside the light cone. As a result, the quantum fields
indeed respect microcausality along the string. The matrices
$\Gamma^\mu$ appear as the analogues of the matrices $\gamma^\mu$ for
the Dirac spinors living in the vortex. The two-dimensional
quantization along the string is thus not independent of the
transverse structure. It is all the more so manifest in the
anticommutator expression between $\Psi$ and $\overline{\Psi}$: from
Eq.~(\ref{reducecom}), and using Eq.~(\ref{gamma03m}), one gets
\begin{eqnarray}
\left\{\Psi({\bold{x}}), \overline{\Psi} ({\bold{x}}^\prime) \right\}
& = & \Gamma^0(\xper,\xper')\left(i\gamma^0 \partial_0 + i\gamma^3
\partial_3 + \M_{\mathrm{d}}(\xper')\right)\, i \Delta(\xpar-\xpar').
\end{eqnarray}
The matrix $\Gamma^0$ now appears clearly as a local transverse
normalization of the longitudinal quantum field operators. Note that
the mass term also depends on the transverse coordinates due to the
nontrivial profile of the Higgs field around the string. Moreover,
setting $t=t'$ in Eq.~(\ref{reducecom}), leads to the postulated
anticommutator at equal times (\ref{psiquantization}), and therefore
justifies the introduction of the $\Gamma^0$ term.

\subsubsection{Fock states}

In the following, $|\Pstate\rangle$ will design a Fock state
constructed by applying creation operators associated with a massive
mode $\miv$, on the relevant string vacuum. Such a state was similarly
defined for zero modes in Ref.~\cite{ringeval}. From the
anticommutators (\ref{creatoranticom}), a massive $\Psi$ state with
momentum $k$ is now normalized according to
\begin{equation}
\langle k'|k \rangle = 2\pi 2\omega \delta(k-k').
\end{equation}
Similarly, it will turn out to be convenient to derive the average of
the occupation number operator since it will appear in the derivation
of the equation of state. From Eq.~(\ref{creatoranticom}), and for a
$\Psi$ massive mode, it reads
\begin{equation}
\label{twoaverage}
\frac{\langle \Pstate |b^\dag(k) b(k')| \Pstate \rangle}{\langle
\Pstate|\Pstate \rangle} = \frac{2 \pi}{L} 2 \omega 2\pi \sum_i
\delta(k-k_i)\delta(k'-k_i),
\end{equation}
where the summation runs over all $\Psi$ massive particle states
present in the relevant $\miv$ Fock state $|\Pstate\rangle$, and $L$
is the physical string length, coming from the $\delta(0)$
regularization by means of Eq.~(\ref{delta}).

\subsection{Stress tensor and Hamiltonian}

The classical stress tensor can immediately be derived from variation
of the full Lagrangian (\ref{lagrangien}) with respect to the metric,
and the $\Psi$ fermionic part thus reads~\cite{ringeval}
\begin{equation}
\label{psitensor}
T_\psi^{\mu \nu} = \frac{i}{2} \overline{\Psi} \gamma^{(\mu}
\partial^{\nu)}
\Psi - \frac{i}{2} \left[\partial^{(\mu}\overline{\Psi}\right]
\gamma^{\nu)} \Psi - B^{(\mu} j^{\nu)}_{_\psi}.
\end{equation}

\subsubsection{Hamiltonian}

The quantum operators associated with the classically conserved
charges can be obtained by replacing the classical fields by their
quantum forms involving creation and annihilation operators. In this
way, the Hamiltonian appears, from Noether theorem, as the charge
associated with the time component of the energy momentum tensor
\begin{equation}
T^{tt}_\psi=i \overline{\Psi}\gamma^0 \partial_t \Psi - i\left(\partial_t
\overline{\Psi} \right) \gamma^0 \Psi.
\end{equation}
Using Eqs.~(\ref{psiuv}) and (\ref{psiexpansion}) in the previous
equation, and performing a spatial integration, the Hamiltonian operator
reads, after some algebra,
\begin{equation}
\displaystyle
P^t_\psi= \int{\frac{\ud k}{2\pi2\omega}}\int{ \ud^2\xper}
\left[-b(k) b^\dag(k) +d^\dag(k) d(k) \right] \overline{u}(k,\xper)
\gamma^0 u(k,\xper).
\end{equation}
In order to simplify this expression, let us introduce the parameters
\begin{eqnarray}
\label{sigmadef}
\Sigmab_{\mathrm{X}}^2  =  \ab_2^2+\ab_3^2, & \quad \textrm{and} \quad &
\Sigmab_{\mathrm{Y}}^2  =  \ab_1^2+\ab_4^2.
\end{eqnarray}
From Eq.~(\ref{psiuv}), the Hamiltonian now reads
\begin{equation}
\label{hamiltonien}
\displaystyle P^t_\psi= \int{\frac{\ud k}{2\pi2\omega}} \left[-b(k)
b^\dag(k) +d^\dag(k) d(k) \right]\left[(\omega -
k)||\Sigma_{\mathrm{X}}^2||+(\omega+k)||\Sigma_{\mathrm{Y}}^2||\right],
\end{equation}
with
\begin{eqnarray}
||\Sigma^2|| = \int{r \, \ud r\, \ud \theta\,\Sigmab^2} = \int{\varrho
\,  \ud \varrho\,\widetilde{\Sigma}^2(\varrho)}.
\end{eqnarray}
Analogous relations also hold for the $\Chi$ field. It is interesting
to note that Eq.~(\ref{hamiltonien}) generalizes the expression
previously derived in the zero modes case~\cite{ringeval}, being found
again by setting $\omega=-k$ for the $\Psi$ zero modes, or $\omega=k$
for the $\Chi$ ones. The normal ordered Hamiltonian is obtained if one
uses the anticommuting normal ordered product for creation and
annihilation operators, i.e.,
\begin{equation}
\label{hamiltonienorder}
\displaystyle
:P^t_\psi:= \int{\frac{\ud k}{2\pi2\omega}}
\left[b^\dag(k)b(k) +d^\dag(k) d(k)
\right]\left[(\omega - k)||\Sigma_{\mathrm{X}}^2||+(\omega+k)||\Sigma_{\mathrm{Y}}^2||\right].
\end{equation}
Since $\omega \ge k$, this Hamiltonian is always positive definite and
is thus well defined. Note that, as in the zero mode case, such a
prescription overlooks the energy density difference between the
vacuum on the string and the usual one, but it can be shown to 
vary as $1/L^2$ and therefore goes to zero in the infinite string
limit~\cite{ringeval,fulling,kay}.

\subsubsection{Effective stress tensor}

In order to make contact with the macroscopic formalism~\cite{formal},
it is necessary to express the classically observable quantities with no
explicit dependence in the microscopic structure. The relevant
two-dimensional fermionic energy momentum tensor can be identified
with the full one in Eq.~(\ref{psitensor}), once the transverse
coordinates have been integrated over. Due to the cylindrical symmetry
around the string direction, $z$ say, all nondiagonal components,
in a Cartesian basis, involving a transverse coordinate vanish after
integration. Moreover, since the fermion fields are normalizable in
the transverse plane, the diagonal terms, $T_\Ferm^{rr}$ and
$T_\Ferm^{\theta\theta}$, are also well defined, and by means of
local transverse stress tensor conservation, the integrated
diagonal components, $T_\Ferm^{xx}$ and $T_\Ferm^{yy}$, also
vanish~\cite{peterpuy}. As expected, the only relevant terms in the
macroscopic formalism are thus $T_\Ferm^{\alpha \beta}$ with $\alpha,
\beta \in \{t,z\}$, i.e., the ones that live only in the string
world sheet. On the other hand, the macroscopic limit of the involved
quantum operators is simply obtained by taking their average over the
relevant Fock states.

Replacing the quantum fields in Eq.~(\ref{psitensor}) by their
expansion (\ref{psiexpansion}), and using Eq.~(\ref{psiuv}), one gets
the quantum expression of the energy momentum tensor. Averaging the
relevant components $T_\psi^{\alpha \beta}$ in the Fock state
$|\Pstate\rangle$, by means of Eqs.~(\ref{psiuv}) and
(\ref{twoaverage}), one obtains
\begin{eqnarray}
\label{ttpsitensor}
\langle :T_\psi^{tt}: \rangle_\Pstate & = & \frac{1}{L}
\left[\sum_i^{N_\psi} \overline{u}(k_i)\gamma^0 u(k_i) +
\sum_j^{\overline{N}_\psi} \overline{u}(k_j)\gamma^0 u(k_j)\right], \\
\label{zzpsitensor}
\langle :T_\psi^{zz}:\rangle_\Pstate & = & \frac{1}{L} \left[
\sum_i^{N_\psi} \frac{k_i}{\omega_i}\overline{u}(k_i)\gamma^3 u(k_i) +
\sum_j^{\overline{N}_\psi}
\frac{k_j}{\omega_j}\overline{u}(k_j)\gamma^3 u(k_j)\right], \\
\label{tzpsitensor}
\langle :T_\psi^{tz}:\rangle_\Pstate & = &
\frac{1}{2L}\left(\sum_i^{N_\psi} \left[\overline{u}(k_i)\gamma^3 u(k_i) +
\frac{k_i}{\omega_i}\overline{u}(k_i)\gamma^0 u(k_i)\right] +
(i,N_\psi)\leftrightarrow (j,\overline{N}_\psi) \right),
\end{eqnarray}
where the  $i$ summations run over the $N_\psi$ particle states
with momentum $k_i$ involved in the Fock state $|\Pstate\rangle$,
while the  $j$ summations take care of the $\overline{N}_\psi$
antiparticle states with momentum $k_j$, all with mass $\miv$. In
order to simplify the notation, the transverse dependence of the unit
spinors have not been written, and the averaged operators stand for
\begin{equation}
\langle T^{\alpha \beta} \rangle_{{\mathcal{P}}} =
\frac{\langle {\mathcal{P}}|T^{\alpha \beta}|{\mathcal{P}}
\rangle}{ \langle {\mathcal{P}}|{\mathcal{P}}\rangle}.
\end{equation}
Similarly, the same relationships can be derived for the $\Chi$ field,
by replacing the $\Psi$ unit spinors by the $\Chi$ ones with the
correct angular dependence, and certainly, in another mass
representation $\miv_\chi$. Once the transverse coordinates have been
integrated over, Eqs.~(\ref{ttpsitensor}), (\ref{zzpsitensor}), and
(\ref{tzpsitensor}), lead to the two-dimensional $\Psi$ stress tensor
\begin{equation}
\label{twopsistress}
\langle \overline{T}_\psi^{\alpha \beta} \rangle_\Pstate =
\left(
\begin{array}{ccc}
\displaystyle E_\psip ||\Sigma_{\mathrm{Y}}^2|| + \overline{E}_\psip
||\Sigma_{\mathrm{X}}^2|| & \displaystyle
\frac{E_\psip+P_\psip}{2}||\Sigma_{\mathrm{Y}}^2||-\frac{\overline{E}_\psip
- \overline{P}_\psip}{2} ||\Sigma_{\mathrm{X}}^2|| \\ \\ \displaystyle
\frac{E_\psip+P_\psip}{2}||\Sigma_{\mathrm{Y}}^2||-\frac{\overline{E}_\psip
- \overline{P}_\psip}{2} ||\Sigma_{\mathrm{X}}^2|| & \displaystyle
P_\psip ||\Sigma_{\mathrm{Y}}^2|| - \overline{P}_\psip
||\Sigma_{\mathrm{X}}^2||
\end{array}
\right),
\end{equation}
with the notations
\begin{equation}
\label{defparam}
\begin{array}{lclclcl}
\displaystyle
E_\psip & = &
\displaystyle\frac{1}{L}
\left[\sum_i^{N_\psi}\left(\omega_i+k_i\right) +
\sum_j^{\overline{N}_\psi} \left(\omega_j+k_j\right)\right],& \quad &
\displaystyle
\overline{E}_\psip & = &
\displaystyle
\frac{1}{L}\left[\sum_i^{N_\psi}\left(\omega_i-k_i\right) +
\sum_j^{\overline{N}_\psi} \left(\omega_j-k_j\right) \right],\\
\displaystyle
P_\psip & = & \displaystyle\frac{1}{L}
\left[\sum_i^{N_\psi}\frac{k_i}{\omega_i}\left(\omega_i+k_i\right) +
\sum_j^{\overline{N}_\psi}
\frac{k_j}{\omega_j}\left(\omega_j+k_j\right)\right],& \quad &
\displaystyle
\overline{P}_\psip & = & \displaystyle\frac{1}{L}
\left[\sum_i^{N_\psi}\frac{k_i}{\omega_i}\left(\omega_i-k_i\right) +
\sum_j^{\overline{N}_\psi}
\frac{k_j}{\omega_j}\left(\omega_j-k_j\right)\right].
\end{array}
\end{equation}
Recall that the full effective energy momentum tensor also involves
the Higgs and gauge fields of the vortex background. Since they
essentially describe a Goto-Nambu string~\cite{gn}, their transverse
integration yields a traceless diagonal tensor
\begin{equation}
\label{gotonambu}
\int{r \,\ud r\,\ud \theta \left(T^{tt}_{\mathrm{g}} +
T^{tt}_{\mathrm{h}} \right)} = -\int{r\,\ud r \,\ud \theta
\left(T^{zz}_{\mathrm{g}} + T^{zz}_{\mathrm{h}} \right)} \equiv M^2.
\end{equation}
Note that the full stress tensor may also involve several massive
$\Psi$ and $\Chi$ states, with different masses belonging to the
spectrum. In this case, there will be as many additional terms in the
form of Eq.~(\ref{twopsistress}), as different massive states there
are in the chosen Fock states.

\subsection{Fermionic currents}

The quantum current operators can be derived from their classical
expressions (\ref{currents}) by using Eq.~(\ref{psiexpansion}), while
the corresponding conserved charges are obtained from their spatial
integration. By means of Eq.~(\ref{twoaverage}), the current operator,
averaged in the relevant Fock state $|\Pstate\rangle$, reads
\begin{eqnarray}
\label{currentaverage}
\langle:j^\alpha_\Ferm:\rangle_\Pstate & = & q \displaystyle{\frac{\cfr+
\cfl}{2}} \frac{1}{L}\left[-\sum_i^{N_\Ferm}
\frac{\overline{u}_{i}\gamma^\alpha u_{i}}{\omega_i} +
\sum_j^{\overline{N}_\Ferm} \frac{\overline{u}_{j}
\gamma^\alpha u_{j}}{\omega_j} \right]  \nonumber
\\
& + & q \displaystyle{\frac{\cfr-
\cfl}{2}} \frac{1}{L}\left[-\sum_i^{N_\Ferm}
\frac{\overline{u}_{i}\gamma^\alpha\gamma_5 u_{i}}{\omega_i} +
\sum_j^{\overline{N}_\Ferm} \frac{\overline{u}_{j}
\gamma^\alpha\gamma_5 u_{j}}{\omega_j} \right],
\end{eqnarray}
with $\alpha \in \{t,z\}$, and once again, the sums run over $\Psi$,
or $\Chi$, particle and antiparticle states. The $u_i$ are the unit
spinors associated with the field dealt with. Concerning the transverse
components, due to the properties of the unit spinors $u$ and $v$ in
Eq.~(\ref{psiuv}), only the orthoradial one does not vanish and reads
\begin{eqnarray}
\langle:j^\theta_\Ferm:\rangle_\Pstate & = &- q \displaystyle{\frac{\cfr+
\cfl}{2}} \frac{1}{L}\left[\sum_i^{N_\Ferm}
\frac{\overline{u}_{i}\gamma^\theta u_{i}}{\omega_i} +
\sum_j^{\overline{N}_\Ferm} \frac{\overline{u}_{j}
\gamma^\theta u_{j}}{\omega_j} \right]  \nonumber
\\
& - & q \displaystyle{\frac{\cfr-
\cfl}{2}} \frac{1}{L}\left[\sum_i^{N_\Ferm}
\frac{\overline{u}_{i}\gamma^\theta\gamma_5 u_{i}}{\omega_i} +
\sum_j^{\overline{N}_\Ferm} \frac{\overline{u}_{j}
\gamma^\theta\gamma_5 u_{j}}{\omega_j} \right],
\end{eqnarray}
whereas $\langle:j_\Ferm^r:\rangle=0$ due to the bound state nature of
the studied currents. As expected, the gauge charges carried by each
trapped fermion in the form of massive mode, generate only macroscopic
charge and current densities along the string, as was the case for the
zero modes~\cite{ringeval}. However, the nonvanishing orthoradial
component shows that the local charges also wind around the string
while propagating in the $z$ direction, as suggested by the above
numerical studies. However, this component will be no longer relevant
in the macroscopic formalism, since it vanishes in a Cartesian basis,
once the transverse coordinates have been integrated over.\\
Nevertheless, this nonzero angular momentum of the massive modes is
found to generate new properties for the longitudinal currents. Let us
focus on the vectorial gauge currents generated by one exitation
state, with energy $\omega$ and momentum $k$, of a $\Psi$
massive mode, $\miv$ say. From Eq.~(\ref{currentaverage}), using
Eqs.~(\ref{psiuv}) and (\ref{sigmadef}), the world sheet vectorial
charge current reads
\begin{eqnarray}
\langle:j^0_{\psi \mathrm{V}}:\rangle_\varepsilon & = &- q
\frac{\cpsir+\cpsil}{2} \frac{\varepsilon}{L}
\left[\left(1+\frac{k}{\omega}\right)\Sigmab_{\mathrm{Y}}^2 + \left(1
- \frac{k}{\omega} \right)\Sigmab_{\mathrm{X}}^2\right],\\
\label{anommoment}
\langle:j^3_{\psi \mathrm{V}}:\rangle_\varepsilon & = &- q
\frac{\cpsir+\cpsil}{2} \frac{\varepsilon}{L}
\left[\left(1+\frac{k}{\omega}\right)\Sigmab_{\mathrm{Y}}^2 - \left(1
- \frac{k}{\omega} \right)\Sigmab_{\mathrm{X}}^2\right],
\end{eqnarray}
where $\varepsilon=\pm 1$ stands a one particle or antiparticle
exitation state. Now, even setting $k=0$ in the previous equations
yields a nonvanishing spatial current. Physically, it can be simply
interpreted as an anomalous magneticlike moment of the considered
massive mode in its rest frame. Examining Eq.~(\ref{anommoment}) shows
that it could be null only if
$\Sigmab_{\mathrm{Y}}^2(r)=\Sigmab_{\mathrm{X}}^2(r)$, which is
generally not satisfied due to the particular shapes of massive spinor
components trapped in the string (see Sec.~\ref{modemassif}). These
ones being associated with nonzero winding numbers, it is therefore
not surprising that, even for a massive stationary state along the
string, the nonvanishing charge angular momentum around the string
generates such additional magneticlike moment. Note that it does not
concern the zero modes, first because they precisely involve
vanishing winding numbers~\cite{ringeval}, and then because for them,
there is no defined rest frame due to their vanishing mass. Obviously,
this property can be generalized for the axial part of the current,
and thus is also valid for the total current of any massive spinor
field trapped in the string.

All the above construction of the Fock space and the derivation of the
quantum operators associated with the energy momentum tensor and gauge
currents remains valid for each $\Psi$ and $\Chi$ massive mode. More
precisely, the other masses belonging to the $\Psi$ spectrum verify
analogous relationships provided $\miv$ is replaced by the relevant
one, as for the unit spinors. In addition, the $\Chi$ massive states
require to transform $n\rightarrow-n$ in Eq.~(\ref{md}), due to their
coupling to the antivortex. At this stage, the averaged values of the
stress tensor and currents have been obtained, and therefore allow the
derivation of an equation of state, once the Fock states are known.

\section{Equation of state}
\label{etat}

The energy per unit length and tension in a given Fock state
$|\Pstate\rangle$ are basically the eigenvalues associated with
timelike and spacelike eigenvectors of the effective two-dimensional
full stress tensor. Obviously this one includes the classical
Goto-Nambu term resulting of the string forming Higgs and gauge fields
[see Eq.~(\ref{gotonambu})], with the fermionic part generated by the
massive currents [see Eq.~(\ref{twopsistress})]. Moreover, in order to
describe the string by an adequate macroscopic
formalism, it is necessary to choose a quantum
statistics for the relevant Fock states, and for energy scales far
below the ones where the string was formed, it is reasonable to
consider a Fermi-Dirac distribution at zero
temperature~\cite{ringeval,prep}. In the following, the equation of
state is first derived for the lowest massive modes associated with
the $\Psi$ and $\Chi$ field, and simplified in the zero-temperature
and infinite string limit. As a second step, these derivations are
generalized to any number and kind of trapped fermionic mode.

\subsection{Lowest massive modes}

\subsubsection{Energy per unit length and tension}

In this section, we will only consider the lowest massive modes
belonging to the $\Psi$ and $\Chi$ mass spectrums, with masses
$\miv_\psi$ and $\miv_\chi$, respectively. From
Eqs.~(\ref{twopsistress}) and (\ref{gotonambu}), the full effective
energy momentum tensor reads
\begin{equation}
\label{lowtwotensor}
\langle\overline{T}^{\alpha \beta}\rangle_\Pstate = \int{r \, \ud r \,
\ud \theta\,\left( T^{\alpha\beta}_{\mathrm{g}} + T^{\alpha
\beta}_{\mathrm{h}}\right)} + \langle\overline{T}^{\alpha
\beta}_\psi\rangle_\Pstate + \langle \overline{T}^{\alpha
\beta}_\chi\rangle_\Pstate,
\end{equation}
where $\overline{T}^{\alpha \beta}_\chi$ takes the same form as
$\overline{T}^{\alpha \beta}_\psi$ in Eqs.~(\ref{twopsistress}) and
(\ref{defparam}) once the $\Psi$ relevant parameters have been
replaced by the $\Chi$ ones. In the preferred frame where the stress
tensor is diagonal, we can identify its timelike and spacelike
eigenvalues with energy per unit length $U$ and tension $T$. Upon
using Eqs.~(\ref{twopsistress}), (\ref{gotonambu}), and
(\ref{lowtwotensor}), these read
\begin{eqnarray}
\label{upsichi}
U_\Pstate & = & M^2+ \sum_{\Ferm \in \{\Psi,\Chi\}}\left[
\frac{E_\Ferm-P_\Ferm}{2}||\Sigma_{\mathrm{Y} \Ferm}^2|| +
\frac{\overline{E}_\Ferm + \overline{P}_\Ferm}{2} ||\Sigma_{\mathrm{X}
\Ferm}^2||\right] \nonumber \\ \vsep & & + \quad \left[\sum_{\Ferm \in
\{\Psi,\Chi\}}\left(E_\Ferm + P_\Ferm\right)||\Sigma_{\mathrm{Y}
\Ferm}^2||\, \times\, \sum_{\Ferm \in
\{\Psi,\Chi\}}\left(\overline{E}_\Ferm - \overline{P}_\Ferm\right)
||\Sigma_{\mathrm{X} \Ferm}^2||\right]^{1/2}\,,
\end{eqnarray}
for the energy per unit length, and
\begin{eqnarray}
\label{tpsichi}
T_\Pstate & = & M^2+ \sum_{\Ferm \in \{\Psi,\Chi\}}\left[
\frac{E_\Ferm-P_\Ferm}{2}||\Sigma_{\mathrm{Y} \Ferm}^2|| +
\frac{\overline{E}_\Ferm + \overline{P}_\Ferm}{2} ||\Sigma_{\mathrm{X}
\Ferm}^2||\right] \nonumber \\ \vsep & & - \quad \left[\sum_{\Ferm \in
\{\Psi,\Chi\}}\left(E_\Ferm + P_\Ferm\right)||\Sigma_{\mathrm{Y}
\Ferm}^2||\,\times\, \sum_{\Ferm \in
\{\Psi,\Chi\}}\left(\overline{E}_\Ferm - \overline{P}_\Ferm\right)
||\Sigma_{\mathrm{X} \Ferm}^2||\right]^{1/2}\,,
\end{eqnarray}
for the tension. It is interesting to note first that
$U_\Pstate+T_\Pstate \neq 2M^2$, and thus the fixed trace equation
of state previously found for zero modes~\cite{ringeval,prep} is no
longer verified by massive modes, as expected since they are no longer
eigenstates of the $\gamma^0\gamma^3$ operator. Moreover, the
expression of energy density and tension does not seem to involve the
conserved charge current magnitude, which played the role of a state
parameter in the case of a scalar condensate in a cosmic
string~\cite{formal,neutral}. In fact, as it was the case at zeroth
order for the zero modes~\cite{ringeval}, the charge currents are only
involved in the stress tensor through their coupling to the gauge
field [see Eq.~(\ref{psitensor})]. At zeroth order, when the back
reaction is neglected, the only nonvanishing component of the gauge
field is $B_\theta$, and it therefore couples only with $j_\Ferm^\theta$,
which vanishes once the transverse coordinates have been integrated
over. As a result, it is not surprising that the equation of state
does not involve the fermionic currents without back reaction. As a
result, it is more natural from quantization to define the occupation
numbers of the involved species as state parameters.

\subsubsection{Zero-temperature and infinite string limit}
\label{zerolimit}
\begin{figure}
\begin{center}
\epsfig{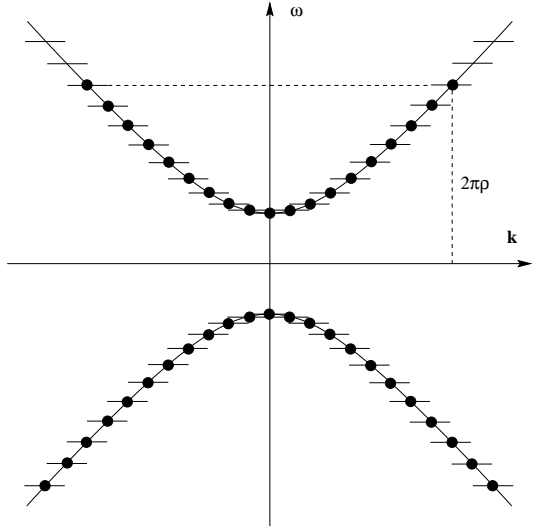}
\caption{The filling of massive states trapped in the string, as
expected in the zero-temperature limit, for a particular mass and for
one species, $\Psi$ or $\Chi$ say. All antiparticles have disappeared
by annihilation with particles during cooling, and the interactions
between particles moving in opposite directions, as their coupling to
the gauge field, lead to the energetically favored configuration with
same number of up and down movers. Obviously, the Fermi level
is necessarily below the vacuum mass of the relevant fermion.}
\label{figfillstates}
\end{center}
\end{figure}

\begin{figure}
\begin{center}
\epsfig{file=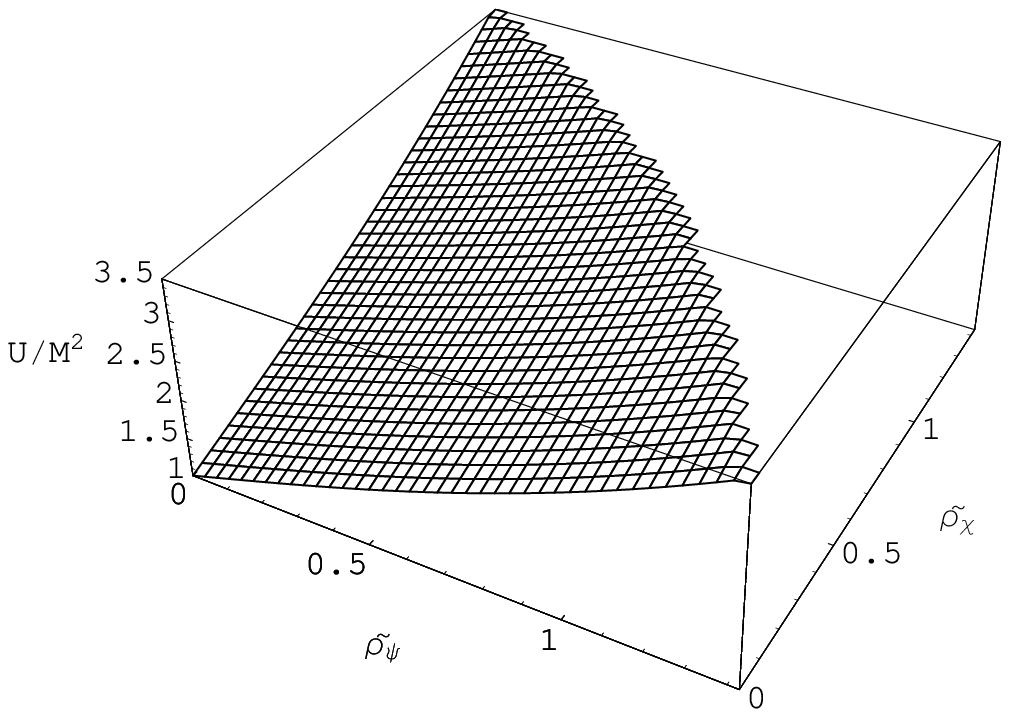,width=8cm}
\epsfig{file=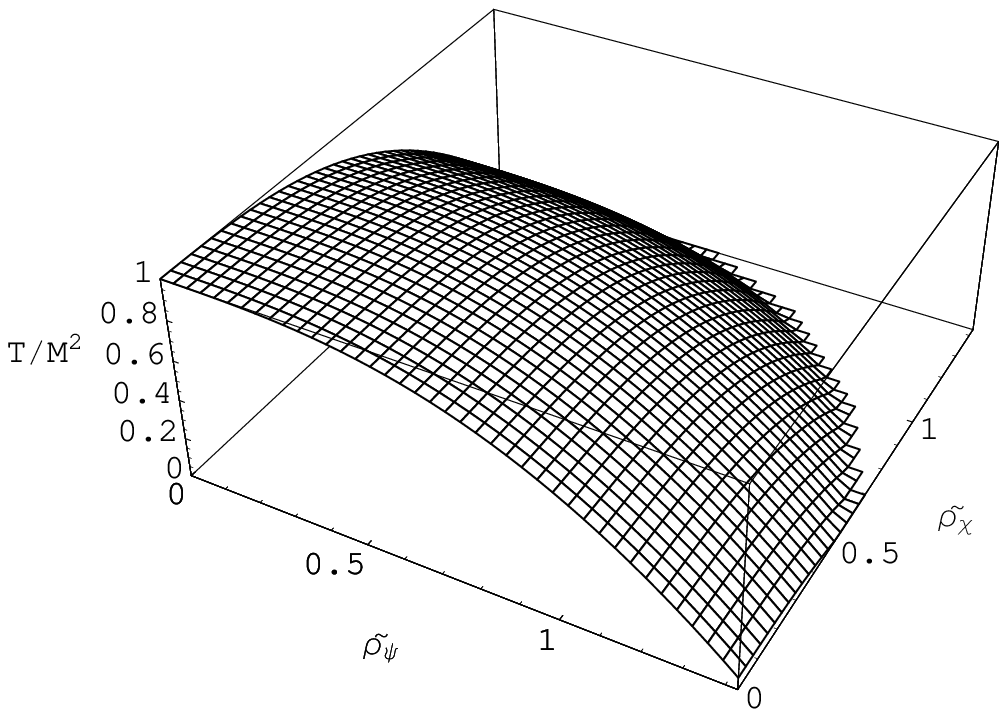,width=8cm}
\caption{The energy per unit length and the tension, in unit of $M^2$,
for the lowest massive modes alone, plotted as function of the
dimensionless effective densities of the two fermion fields,
$\rhot_\psi$ and $\rhot_\chi$. The parameters have been chosen in the
nonperturbative sector, with $m_{\mathrm{f}}/m_{\mathrm{h}}\sim 3$
and $\miv_\chi \sim \miv_\psi \sim 0.6 m_{\mathrm{f}}$. Note the
linear variation of the energy density near the origin whereas the
tension varies quadratically. Moreover, in the allowed range for
fermion densities, i.e., less than the fermion vacuum mass, the
tension vanishes and the string becomes unstable with respect to
transverse perturbations.}
\label{figutm1}
\end{center}
\end{figure}

Assuming a Fermi-Dirac distribution at zero temperature for the
exitation states, the sums involved in Eq.~(\ref{defparam}) run over
the successive values of the allowed momentum $k_i$ until the Fermi
level of the considered species is reached. With periodic boundary
conditions on spinor fields, the allowed momentum exitation values are
discretized according to
\begin{equation}
k_n=\frac{2\pi}{L}n,
\end{equation}
where $n$ is an integer, playing the role of a quantum exitation
number. As a result, in the relevant $\miv$ representation of each
field, the exitation energies $\omega_i$ are also discrete according
to Eq.~(\ref{dispmass}), and for the $\Psi$ field, the parameters
$E_\psi$ and $P_\psi$ in Eq.~(\ref{defparam}) simplify to sums of
radical function of $n$, with $n$ running from the vacuum to the last
filled state. In order to express them as explicit functions of the
relevant Fermi level, it is convenient to consider the infinite
string limit $L\rightarrow\infty$. In this limit one gets
\begin{equation}
\label{thermlimit}
\lim_{L \to \infty}\frac{1}{L}\sum_{i=-N_\psi^-}^{N_\psi^+}f(k_i) =
\frac{1}{2\pi} \int_{-2 \pi \rho_\psi^-} ^{2\pi\rho_\psi^+} \ud k\,f(k),
\end{equation}
where $\rho_\psi^\pm=N_\psi^\pm/L$ are the $\Psi$ up and down
mover densities, $N_\psi^+$, $N_\psi^-$ standing for the number of
$\Psi$ particle moving in the $+z$ or $-z$ directions,
respectively. Note that the total number of particles of this
kind is thus $N_\psi=N_\psi^++N_\psi^-+1$ since there is the
additional rest state obtained for $k=0$. After some algebra in
Eq.~(\ref{defparam}), using Eq.~(\ref{thermlimit}), the parameters,
for the $\Psi$ field, read
\begin{eqnarray}
\label{limitparam}
E_\psi & = &
\frac{\miv^2}{4\pi}\left(\rhot_\psi^{+^2}-\rhot_\psi^{-^2} +
\left[\rhot_\psi^+ \sqrt{1+\rhot_\psi^{+^2}} + \rhot_\psi^- \sqrt{1 +
\rhot_\psi^{-^2}}\right] + \ln{\left[\left(\sqrt{1 + \rhot_\psi^{+^2}}
+ \rhot_\psi^{+}\right)\left(\sqrt{1 + \rhot_\psi^{-^2}} +
\rhot_\psi^{-}\right)\right]} \right) \nonumber \\ & & +
\left(\rhot_\psi^\pm \leftrightarrow \overline{\rhot}_\psi^\pm\right),
\\ P_\psi & = &
\frac{\miv^2}{4\pi}\left(\rhot_\psi^{+^2}-\rhot_\psi^{-^2} +
\left[\rhot_\psi^+ \sqrt{1+\rhot_\psi^{+^2}} + \rhot_\psi^- \sqrt{1 +
\rhot_\psi^{-^2}}\right] - \ln{\left[\left(\sqrt{1 + \rhot_\psi^{+^2}}
+ \rhot_\psi^{+}\right)\left(\sqrt{1 + \rhot_\psi^{-^2}} +
\rhot_\psi^{-}\right)\right]} \right) \nonumber \\ & & +
\left(\rhot_\psi^\pm \leftrightarrow \overline{\rhot}_\psi^\pm\right),
\end{eqnarray}
where $\rhot_\psi$ stands for the dimensionless $\Psi$ mover density
\begin{equation}
\label{adimdens}
\rhot_\psi=\frac{2\pi}{\miv}\rho_\psi,
\end{equation}
while $\overline{\rhot}_\psi$ is defined in the same way for the
$\Psi$ antiparticle states. Similarly, the two other parameters
$\overline{E}_\psi$ and $\overline{P}_\psi$ read
\begin{eqnarray}
\overline{E}_\psi & = &
\frac{\miv^2}{4\pi}\left(-\rhot_\psi^{+^2}+\rhot_\psi^{-^2} +
\left[\rhot_\psi^+ \sqrt{1+\rhot_\psi^{+^2}} + \rhot_\psi^- \sqrt{1 +
\rhot_\psi^{-^2}}\right] + \ln{\left[\left(\sqrt{1 + \rhot_\psi^{+^2}}
+ \rhot_\psi^{+}\right)\left(\sqrt{1 + \rhot_\psi^{-^2}} +
\rhot_\psi^{-}\right)\right]} \right) \nonumber \\ & & +
\left(\rhot_\psi^\pm \leftrightarrow \overline{\rhot}_\psi^\pm\right),
\\
\label{limitparambar}
\overline{P}_\psi & = &
\frac{\miv^2}{4\pi}\left(\rhot_\psi^{+^2}-\rhot_\psi^{-^2} -
\left[\rhot_\psi^+ \sqrt{1+\rhot_\psi^{+^2}} + \rhot_\psi^- \sqrt{1 +
\rhot_\psi^{-^2}}\right] + \ln{\left[\left(\sqrt{1 + \rhot_\psi^{+^2}}
+ \rhot_\psi^{+}\right)\left(\sqrt{1 + \rhot_\psi^{-^2}} +
\rhot_\psi^{-}\right)\right]} \right) \nonumber \\ & & +
\left(\rhot_\psi^\pm \leftrightarrow \overline{\rhot}_\psi^\pm\right).
\end{eqnarray}
Note that these parameters depend differently on the up and
down mover densities as expected for chiral coupling of the
fermions to the string forming Higgs field. Recall that in the
massless case the zero modes associated with the $\Psi$ and $\Chi$
fields can only propagate in the $-z$ and $+z$ direction
respectively~\cite{witten,ringeval,jackiwrossi}. The same
relationships also hold for the $\Chi$ field by using the relevant
dimensionless mover densities $\rhot^\pm_\chi$ and
$\overline{\rhot}_\chi^\pm$. Although the equation of state can be
derived as a function of these four parameters for each fermion field
$\Ferm$, it is convenient at this stage to perform some physical
simplifications. Contrary to the zero mode case, the coupling between
massive particles and antiparticles of the same species $\Ferm$ does
not vanish along the string. As a result, it is reasonable to consider
that the only kind surviving at zero temperature corresponds to the
one which was in excess in the plasma in which the string was formed
during the phase transition. On the other hand, the energetically
favored distribution at zero temperature involves necessarily the same
number of $\Ferm$ ``up'' and ``down'' movers, each filling the
accessible states living on each branch of the mass hyperbola (see
Fig.~\ref{figfillstates}). As a result, in the considered energy
scale, it seems reasonable to consider only one state parameter per
mass instead of the four initially introduced by quantization, namely
$\rhot_\Ferm=\rhot_\Ferm^+=\rhot_\Ferm^-$, for a plasma dominated by
particles, say.
Setting these simplifications in Eqs.~(\ref{limitparam}) to
(\ref{limitparambar}), by means of Eqs.~(\ref{upsichi}) and
(\ref{tpsichi}) the energy density and the tension associated with the
lowest massive modes now read
\begin{eqnarray}
\label{loweqstate}
U & = & M^2+ \frac{1}{2\pi}\sum_{\Ferm} \miv_\Ferm^2 \ln{\left[\sqrt{1
+ \rhot_\Ferm^2} + \rhot_\Ferm\right]} \nonumber \\ & & +
\frac{1}{\pi}\,\left[\sum_\Ferm \miv_\Ferm^2 \, ||\Sigma_{\mathrm{Y}
\Ferm}^2||\, \rhot_\Ferm \sqrt{1+\rhot_\Ferm^2}\,\times\, \sum_\Ferm
\miv_\Ferm^2 \, ||\Sigma_{\mathrm{X} \Ferm}^2||\, \rhot_\Ferm
\sqrt{1+\rhot_\Ferm^2}\right]^{1/2}\,,\\ T & = & M^2+
\frac{1}{2\pi}\sum_{\Ferm} \miv_\Ferm^2 \ln{\left[\sqrt{1 +
\rhot_\Ferm^2} + \rhot_\Ferm \right]} \nonumber \\ & & -
\frac{1}{\pi}\,\left[\sum_\Ferm \miv_\Ferm^2 \, ||\Sigma_{\mathrm{Y}
\Ferm}^2||\, \rhot_\Ferm \sqrt{1+\rhot_\Ferm^2}\,\times\, \sum_\Ferm
\miv_\Ferm^2 \, ||\Sigma_{\mathrm{X} \Ferm}^2||\, \rhot_\Ferm
\sqrt{1+\rhot_\Ferm^2}\right]^{1/2}\,.
\end{eqnarray}
The sum runs over the two lowest massive bound states, each one being
associated to the two fermion fields trapped in the vortex, namely
$\Psi$ and $\Chi$, and have $\miv_\psi$ and $\miv_\chi$ masses,
respectively. As a result, the equation of state involves two
independent parameters, $\rhot_\psi$ and $\rhot_\chi$, in the
zero-temperature and infinite string limit. The energy per unit length
and the tension have been plotted in Fig.~\ref{figutm1}, for the
lowest massive modes in the nonperturbative sector. The curves are
essentially the same in the perturbative sector, but the variations
around the Goto-Nambu case $U=T=M^2$ are much more damped. For
reasonable values of the transverse normalizations,
e.g., $||\Sigma_{\mathrm{X}}^2|| \sim ||\Sigma_{\mathrm{Y}}^2|| \sim
0.5$, and for small values of the dimensionless parameters
$\rhot_\psi$ and $\rhot_\chi$, the energy density is found to grow
linearly with $\rhot_\psi$ and $\rhot_\chi$, whereas the tension
varies quadratically. As can be seen in Eq.~(\ref{loweqstate}), due to
the minus sign in $T$, all linear terms in $\rhot$ vanish near origin,
whereas it is not the case for the energy density. However, for higher
values of the densities, the quadratic terms dominate and both energy
density and tension end up being quadratic functions of $\rhot$. On
the other hand, according to the macroscopic formalism~\cite{formal},
the string becomes unstable with respect to transverse perturbations
when the tension takes on negative values, as in Fig.~\ref{figutm1}
for high densities. Moreover, the decrease of the tension is more
damped in the perturbative sector, and the negative values cannot
actually be reached for acceptable values of $\rhot$, i.e., $\rhot <
m_{\mathrm{f}}/\miv$. As a result, the rapid decrease of the tension
with respect to the fermion densities constrains the nonperturbative
sector where the string is able to carry massive fermionic
currents. For each mass, the higher acceptable value of the $\rhot$
ensuring transverse normalizability is roughly $m_{\mathrm{f}}/\miv$,
and from Eq.~(\ref{loweqstate}), the tension becomes negative at this
density for $m_{\mathrm{f}}^2 \sim 4\pi M^2$. Much higher values of
$m_{\mathrm{f}}$ will thus yield to empty massive states.
As previously noted, the energy density and tension for massive modes
no longer verify the fixed trace equation of state found with the zero
modes alone. As a result, the longitudinal perturbation propagation
speed $c_{\mathrm{L}}^2=-dT/dU$ is no longer equal to the speed of
light, and it is even no longer well defined since the equation of
state involves more than one state parameter. A necessary condition
for longitudinal stability can nevertheless be stated by verifying
that all the perturbation propagation speeds $-(\partial
T/\partial\rhot)/(\partial U/\partial\rhot)$ obtained from variation
of only one state parameter are positive and less than the speed of
light. The longitudinal and transverse perturbation propagation
speeds, $c^2_{\mathrm{L}}$ and $c^2_{\mathrm{T}}=T/U$, respectively,
have been plotted in Fig.~\ref{figclctm1} in the case where there is
only one species trapped as massive mode, $\Psi$ or $\Chi$ say. It is
interesting to note that there is a transition between a supersonic
regime obtained at low fermion density, and a subsonic at high fermion
density. Moreover, the transition density between the two regimes is
all the more so high as the coupling constant
$m_{\mathrm{f}}/m_{\mathrm{h}}$ is weak. It is not surprising to
recover such zero-mode-like subsonic behavior~\cite{ringeval,prep} for
densities much higher than the rest mass, since in these cases the
ultrarelativistic limit applies. On the other hand, since the mass of
the massive mode decreases with the coupling constant as in
Fig.~{\ref{figmasspect}, the transition will occur earlier in the
nonperturbative sector, as can be seen in
Fig.~\ref{figrhotcross}. Note that the subsonic region is also limited
by the maximum allowed values of the massive fermion densities, i.e,
$\sim m_{\mathrm{f}}/\miv$, and the regions of transverse
instabilities where $c^2_{\mathrm{T}}$ becomes negative. Inclusion of
the other species does not change significantly these behaviors, the
main effect being to lower $c_{\mathrm{T}}^2$ with respect to the
other fermion density, as can be seen in Fig.~\ref{figutm1}.
\begin{figure}
\begin{center}
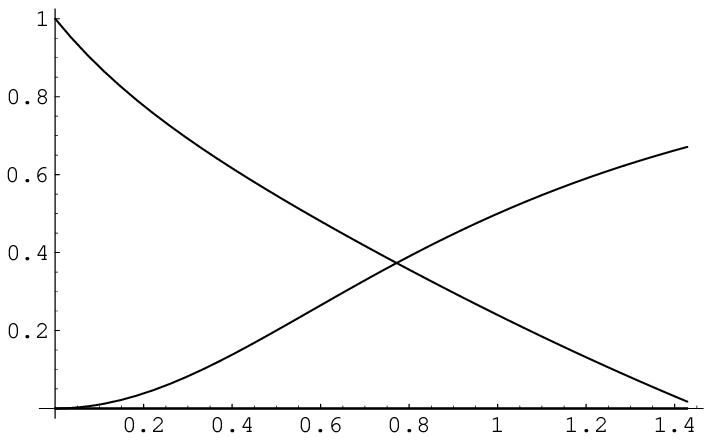
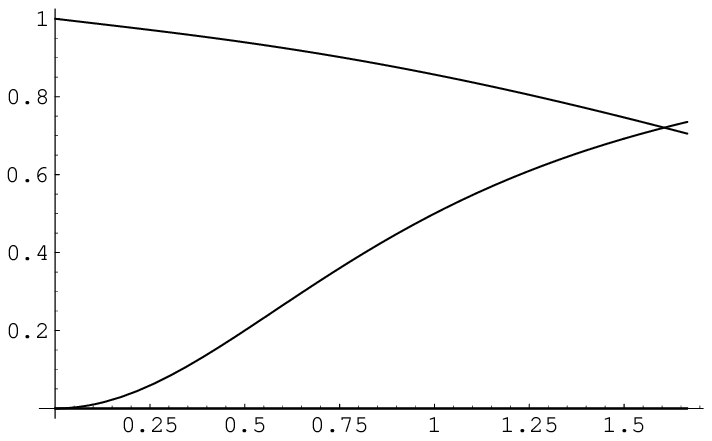
\caption{The squared longitudinal and transverse perturbations
propagation speeds for one massive species only, plotted as functions
of the dimensionless fermion density $\rhot$, for two values of the
coupling constant $m_{\mathrm{f}}/m_{\mathrm{h}}$. Note the transition
between subsonic and supersonic behaviors takes place at a cross
density, $\rhot_\times$ say, which decreases with the coupling
constant (see Fig.~\ref{figrhotcross}).}
\label{figclctm1}
\end{center}
\end{figure}
\begin{figure}
\begin{center}
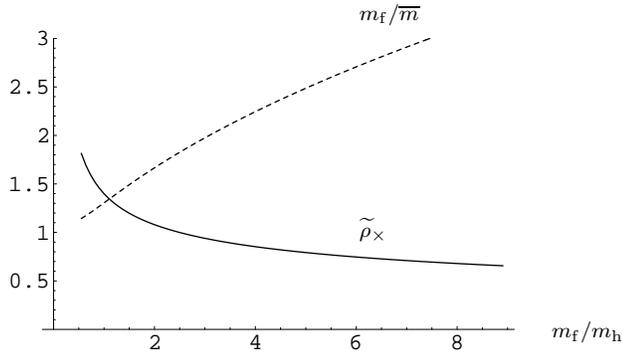
\caption{The cross dimensionless density $\rhot_\times$, i.e., the
dimensionless fermion density for which the transverse and
longitudinal perturbation propagation speeds are equal, plotted as
function of the coupling constant $m_{\mathrm{f}}/m_{\mathrm{h}}$, for
one massive species only. The dashed curve shows the maximum allowed
values $m_{\mathrm{f}}/\miv$ of the massive fermion density ensuring
transverse normalizability. The transition from supersonic regime to
subsonic can thus occurs only in the nonperturbative sector, below
this frontier.}
\label{figrhotcross}
\end{center}
\end{figure}
The current magnitude can also be derived from the averaged current
operators in the zero temperature and infinite string limit. From
Eq.~(\ref{currentaverage}), using Eqs.~(\ref{psiuv}) and
(\ref{twoaverage}), once the transverse coordinates have been
integrated over, the world sheet components read
\begin{equation}
\label{gaugecurrents}
\begin{array}{lcl}
\vsep \displaystyle \langle\overline{j}^0\rangle & = & \displaystyle -
\sum_\Ferm \frac{\miv_\Ferm}{\pi} \left(q \cfr ||\ab_1^2+\ab_2^2|| + q
\cfl ||\ab_4^2 + \ab_3^2|| \right) \,\rhot_\Ferm, \\ \vsep
\displaystyle \langle\overline{j}^3\rangle & = & \displaystyle
-\sum_\Ferm \frac{\miv_\Ferm}{\pi} \left(q \cfr ||\ab_1^2-\ab_2^2|| +
q \cfl ||\ab_4^2 - \ab_3^2|| \right)\,\rhot_\Ferm.
\end{array}
\end{equation}
The current magnitude
${\mathcal{C}}^2 = \langle \overline{j}^0\rangle^2 -
\langle\overline{j}^3 \rangle^2$
therefore reads
\begin{eqnarray}
\label{currentmag}
{\mathcal{C}}^2 & = & 4 \left(\sum_\Ferm \frac{\miv_\Ferm}{\pi}
F_{\mathrm{Y} \Ferm}\, \rhot_\Ferm\right) \left(\sum_\Ferm
\frac{\miv_\Ferm}{\pi} F_{\mathrm{X} \Ferm}\, \rhot_\Ferm\right),
\end{eqnarray}
where $F_{\mathrm{X} \Ferm}$ and $F_{\mathrm{Y} \Ferm}$ denote the
transverse effective charges
\begin{equation}
\begin{array}{lcl}
\vsep \displaystyle F_{\mathrm{Y} \Ferm} & = & \displaystyle q \cfr
||\ab_1^2|| + q \cfl ||\ab_4^2|| \\ \vsep \displaystyle F_{\mathrm{X}
\Ferm} & = & \displaystyle q \cfr ||\ab_2^2|| + q \cfl ||\ab_3^2||,
\end{array}
\end{equation}
already introduced for the zero mode currents~\cite{ringeval}.
In the case of one massive species only, the charge current magnitude
simplifies to
\begin{equation}
{\mathcal{C}}^2=4 \frac{\miv^2}{\pi^2} F_{\mathrm{X}} F_{\mathrm{Y}}
\rhot^2,
\end{equation}
and thus the sign of ${\mathcal{C}}^2$ is only given by the sign of
$F_{\mathrm{X}} F_{\mathrm{Y}}$, which is generally positive for
reasonable values of the transverse normalizations,
e.g., $||\Sigma_{\mathrm{X}}^2|| \sim ||\Sigma_{\mathrm{Y}}^2|| \sim
0.5$. As a result, the charge current generated by only one massive
species is always timelike~\cite{davisC}, contrary to the zero mode
charge current which was found to be possibly timelike, but also
spacelike~\cite{ringeval}, owing to the allowed exitations of
antiparticle zero mode states. As noted above, the antiparticle states
cannot exist for massive modes due to the nonvanishing cross section
along the string between massive particles and antiparticles.
Moreover, and as it was the case for the zero modes, unless there is
only one massive species trapped in the string, the magnitude of the
charge current is not a sufficient state parameter, contrary to the
bosonic current-carrier case~\cite{neutral}.

\subsection{General case}

From the numerical approach in Sec.~(\ref{numresults}), as soon as
the nonperturbative sectors are considered, additional massive bound
states become relevant, and it is reasonable to consider that, in the
zero-temperature limit, all these accessible massive states will be
also be filled. Moreover, the complete description of the string state
also requires the inclusion of the zero modes in addition to the
massive ones.

\subsubsection{Full stress tensor}

The effective two-dimensional energy momentum tensor, involving all
trapped modes in the cosmic string, can be obtained from
Eq.~(\ref{lowtwotensor}) by replacing the sum over the two lowest
massive modes with the sum over all the accessible masses, plus the
zero mode terms previously derived in Ref.~\cite{ringeval}. In the
preferred frame where the stress tensor is diagonal, after some
algebra, the energy density and tension therefore read
\begin{eqnarray}
\label{ueqstate}
U & = & M^2+ \frac{1}{2\pi}\sum_{\Ferm, \ell} \miv_{\Ferm_\ell}^2
\ln{\left[\sqrt{1 + \rhot_{\Ferm_\ell}^2} + \rhot_{\Ferm_\ell}\right]}
\nonumber \\ & & + \frac{1}{\pi}\,\left[\left(4\pi^2 \Rho_{\chi}^2 +
\sum_{\Ferm, \ell} \miv_{\Ferm_\ell}^2 \, ||\Sigma_{\mathrm{Y}
\Ferm_\ell}^2||\, \rhot_{\Ferm_\ell}
\sqrt{1+\rhot_{\Ferm_\ell}^2}\right)\left(4\pi^2 \Rho_{\psi}^2 +
\sum_{\Ferm, \ell} \miv_{\Ferm_\ell}^2 \, ||\Sigma_{\mathrm{X}
\Ferm_\ell}^2||\, \rhot_{\Ferm_\ell}
\sqrt{1+\rhot_{\Ferm_\ell}^2}\right)\right]^{1/2}\,,\\
\label{teqstate}
T & = & M^2+ \frac{1}{2\pi}\sum_{\Ferm, \ell} \miv_{\Ferm_\ell}^2
\ln{\left[\sqrt{1 + \rhot_{\Ferm_\ell}^2} + \rhot_\Ferm \right]}
\nonumber \\ & & - \frac{1}{\pi}\,\left[\left(4\pi^2 \Rho_{\chi}^2 +
\sum_{\Ferm, \ell} \miv_{\Ferm_\ell}^2 \, ||\Sigma_{\mathrm{Y}
\Ferm_\ell}^2||\, \rhot_{\Ferm_\ell}
\sqrt{1+\rhot_{\Ferm_\ell}^2}\right)\left(4\pi^2 \Rho_{\psi}^2 +
\sum_{\Ferm, \ell} \miv_{\Ferm_\ell}^2 \, ||\Sigma_{\mathrm{X}
\Ferm_\ell}^2||\, \rhot_{\Ferm_\ell}
\sqrt{1+\rhot_{\Ferm_\ell}^2}\right)\right]^{1/2}\,.
\end{eqnarray}
The sums run over all accessible massive bound states $\ell$
with masses $\miv_{\Ferm_\ell}$ of each fermion $\Ferm$, i.e., $\Psi$
and $\Chi$. The additional parameters $\Rho_\chi$ and $\Rho_\psi$ are
the particle densities trapped in the string in the form of zero modes,
for the $\Chi$ and $\Psi$ field, respectively, with same notation as
in Ref.~\cite{ringeval}. Note that the zero mode contribution can also be
obtained from the null mass limit in Eq.~(\ref{loweqstate}). As a
result, the full expression of energy per unit length and tension
seems to involve as many state parameters as trapped modes in the
string.

\subsubsection{Equation of state}

As for the lowest massive modes, it is convenient to perform some
approximations owing to the energetically favored filling of the
involved states, in the zero-temperature limit. In particular, it is
reasonable to consider that the nonvanishing cross sections between
massive modes, and between zero modes and massive modes, lead to the
filling of all the accessible states with energy lower than a Fermi
energy, $\Eferm_\Ferm$ say, for each fermion field $\Ferm$. As a
result, the energetically favored filling takes place by successive
jumps from the lower masses to the highest ones, until the last mass
hyperbola with $\miv_{\Ferm_\ell} \sim \Eferm_\Ferm$ is
reached. Obviously, this filling begins with the zero modes, next with
the lowest massive modes and so on. On the other hand, only the
particle states are assumed to be relevant because of the assumed
annihilation of the antiparticle states, as discussed in
Sec.~(\ref{zerolimit}). As a result, the Fermi levels, $\nu_\Ferm$
say, can be defined through the zero modes filling only, as the line
densities of zero mode exitations trapped in the string (see
Fig.~\ref{figallfilled}), and thus play the role of state
parameters.
\begin{figure}
\begin{center}
\epsfig{file=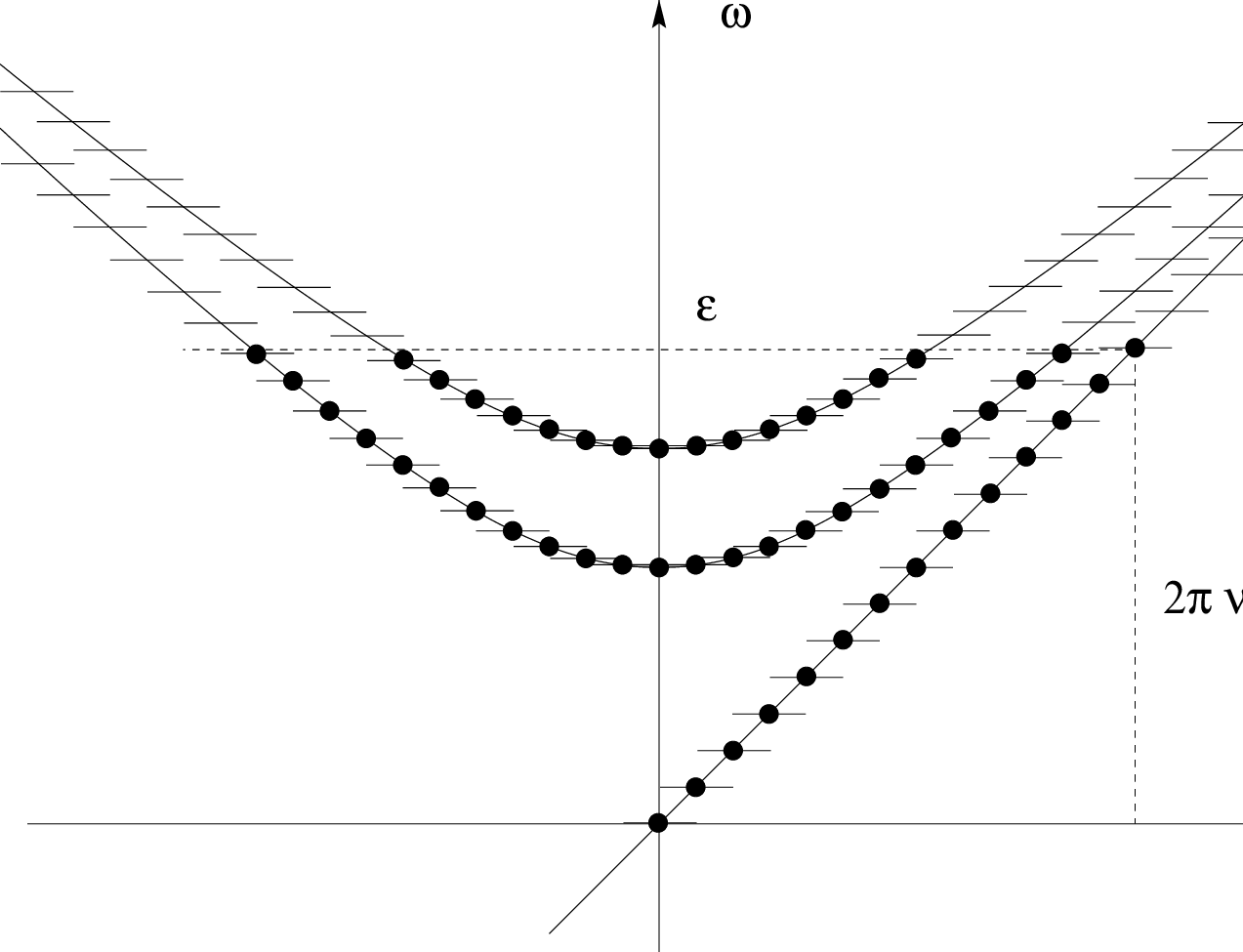,width=8cm}
\caption{The accessible states, for the $\Chi$ fermions, in the
zero-temperature limit. The zero modes are represented by the chiral
line $\omega=k$, while the massive modes appear as mass
hyperbolae. The Fermi levels are therefore dependent of the considered
energy scale $\Eferm$, since the filling is performed by successive
jumps from the zero modes to the massive ones, with $\miv_{\ell} \sim
\Eferm$. As a result, under these approximations, each trapped species
leads to only one state parameter which can be identified with the
Fermi level of the zero mode exitations, namely
$\nu=\Eferm/2\pi$. Note that the antiparticle states have not been
represented due to their assumed annihilation.}
\label{figallfilled}
\end{center}
\end{figure}
According to the so-defined state parameters, the massive
exitation densities $\rho_{\Ferm_\ell}$, in Eqs.~(\ref{ueqstate}) and
(\ref{teqstate}), reduce to
\begin{eqnarray}
\label{massdens}
\rho_{\Ferm_\ell} & = & 
\left(\nu_\Ferm - \frac{\miv_{\Ferm_\ell}}{2\pi} \right) \Theta
\left[\nu_\Ferm - \frac{\miv_{\Ferm_\ell}}{2\pi}\right],
\end{eqnarray}
with $\Theta$ function is the Heavyside step function, as expected for
energy scales less than the rest mass of the considered massive mode.
The zero mode density simply reads
\begin{eqnarray}
\label{zerodens}
\Rho_\Ferm & = & \nu_\Ferm,
\end{eqnarray}
for zero mode particle states alone. From Eqs.~(\ref{massdens}) and
(\ref{zerodens}), and the definition of the dimensionless densities in
Eq.~(\ref{adimdens}),
\begin{equation}
\rhot_{\Ferm_\ell}=\frac{2 \pi}{\miv_{\Ferm_\ell}} \rho_{\Ferm_\ell},
\end{equation}
the energy per unit length and the tension now depend explicitly of
the two state parameters only, namely $\nu_\psi$ and $\nu_\chi$. By
means of Eq.~(\ref{ueqstate}), the energy density reads
\begin{eqnarray}
\label{fullenergy}
U & = & \displaystyle M^2+ \frac{1}{2\pi}\sum_{\miv_{\Ferm_\ell} \le 2
\pi \nu_\Ferm} \miv_{\Ferm_\ell}^2 \ln{\left[\sqrt{1 +
\left(\frac{2\pi}{\miv_{\Ferm_\ell}}\nu_\Ferm-1\right)^2} +
\frac{2\pi}{\miv_{\Ferm_\ell}}\nu_\Ferm-1 \right]} \nonumber \\ & & +
\, \frac{1}{\pi}\,\left[(2 \pi \nu_\chi)^2 + \sum_{\miv_{\Ferm_\ell}
\le 2 \pi \nu_\Ferm} \miv_{\Ferm_\ell}^2 \, ||\Sigma_{\mathrm{Y}
\Ferm_\ell}^2||\,\left(\frac{2\pi}{\miv_{\Ferm_\ell}}\nu_\Ferm-1
\right) \sqrt{1+\left(\frac{2\pi}{\miv_{\Ferm_\ell}} \nu_\Ferm- 1
\right)^2}\right]^{1/2} \nonumber \\ & & \times \, \left[(2 \pi
\nu_\psi)^2 + \sum_{\miv_{\Ferm_\ell} \le 2 \pi \nu_\Ferm}
\miv_{\Ferm_\ell}^2 \, ||\Sigma_{\mathrm{X}
\Ferm_\ell}^2||\,\left(\frac{2\pi}{\miv_{\Ferm_\ell}}\nu_\Ferm-1
\right) \sqrt{1+\left(\frac{2\pi}{\miv_{\Ferm_\ell}} \nu_\Ferm- 1
\right)^2}\right]^{1/2}\,,
\end{eqnarray}
while the tension is obtained from Eq.~(\ref{teqstate}),
\begin{eqnarray}
\label{fulltension}
T & = & \displaystyle M^2+ \frac{1}{2\pi}\sum_{\miv_{\Ferm_\ell} \le 2
\pi \nu_\Ferm} \miv_{\Ferm_\ell}^2 \ln{\left[\sqrt{1 +
\left(\frac{2\pi}{\miv_{\Ferm_\ell}}\nu_\Ferm-1\right)^2} +
\frac{2\pi}{\miv_{\Ferm_\ell}}\nu_\Ferm-1 \right]} \nonumber \\ & & - \,
\frac{1}{\pi}\,\left[(2 \pi \nu_\chi)^2 + \sum_{\miv_{\Ferm_\ell} \le 2
\pi \nu_\Ferm} \miv_{\Ferm_\ell}^2 \, ||\Sigma_{\mathrm{Y}
\Ferm_\ell}^2||\,\left(\frac{2\pi}{\miv_{\Ferm_\ell}}\nu_\Ferm-1
\right) \sqrt{1+\left(\frac{2\pi}{\miv_{\Ferm_\ell}} \nu_\Ferm- 1
\right)^2}\right]^{1/2}  \nonumber \\ & & \times \,
\left[(2 \pi \nu_\psi)^2 + \sum_{\miv_{\Ferm_\ell} \le 2
\pi \nu_\Ferm} \miv_{\Ferm_\ell}^2 \, ||\Sigma_{\mathrm{X}
\Ferm_\ell}^2||\,\left(\frac{2\pi}{\miv_{\Ferm_\ell}}\nu_\Ferm-1
\right) \sqrt{1+\left(\frac{2\pi}{\miv_{\Ferm_\ell}} \nu_\Ferm- 1
\right)^2}\right]^{1/2}\,.
\end{eqnarray}
The full energy per unit length and tension have been plotted in
Fig.~{\ref{figuttwomass} for a configuration including two massive
bound states, in addition to the zero mode ones. Due to the
zero-temperature limit, for densities smaller than the first
accessible mass, the Heavyside functions in Eq.~(\ref{massdens})
vanish, as a result, from Eqs.~(\ref{fullenergy}) and
(\ref{fulltension}) the fixed trace equation of state is
recovered~\cite{ringeval} with
\begin{equation}
\begin{array}{lclclcl}
\displaystyle
U & = &
\displaystyle
M^2 + 4 \pi \nu_\chi \nu_\psi,
& \quad &
\displaystyle
T & = &
\displaystyle
M^2 - 4 \pi \nu_\chi \nu_\psi.
\end{array}
\end{equation}
Once the first mass hyperbola is reached, the behaviors of the energy
per unit length and tension are clearly modified and become very
rapidly dominated by the mass terms, and, as found for the lowest
massive modes alone, the energy density begins to grow linearly with
respect to the state parameters, whereas the tension decreases
quadratically. Actually, the plotted curves in
Fig.~\ref{figuttwomass} show slope discontinuities each time the
phase space is enlarged due to the income of accessible massive bound
states (see Fig.~\ref{figallfilled}).
\begin{figure}
\begin{center}
\epsfig{file=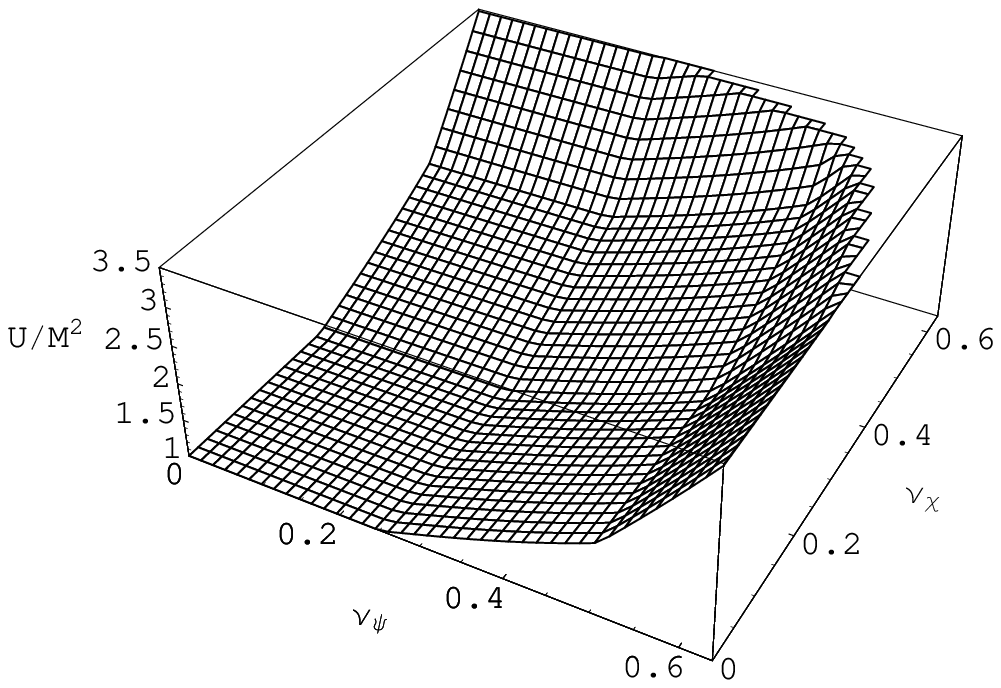,width=8cm}
\epsfig{file=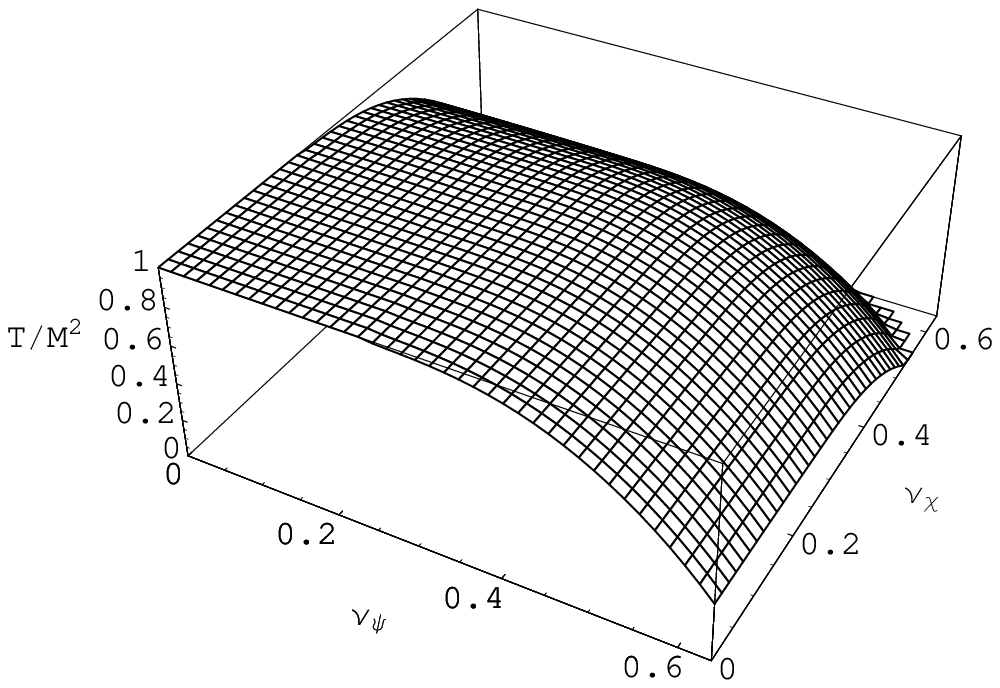,width=8cm}
\caption{The energy per unit length and the tension plotted as
function of the two state parameters, i.e., the zero mode densities,
in the nonperturbative sector $m_{\mathrm{f}}/m_{\mathrm{h}} \sim
4$. Two additional massive bound states have been considered with
respective masses $\miv/m_{\mathrm{f}} \sim 0.4$ and
$\miv/m_{\mathrm{f}} \sim 0.6$. In the zero-temperature limit, the
filling of the accessible states is performed by successive jumps as
soon as the Fermi level reaches one mass hyperbola (see
Fig.~\ref{figallfilled}). As a result, for the lowest values of the
state parameters, only the zero modes are relevant and the fixed
trace equation of state, $U+T=2M^2$, is verified, then the first and
second massive modes are successively reached and become rapidly
dominant. As can be seen near the origin, the smooth variations
induced by the zero modes appear completely negligible compared to the
massive ones. In the perturbative sectors, these behaviors are
essentially the same, but the induced variations of the density
energy and the tension are all the more small.}
\label{figuttwomass}
\end{center}
\end{figure}
On the other hand, it is reasonable to expect competition between the
subsonic regimes induced by zero mode currents, or ultrarelativistic
massive modes, and the supersonic ones coming from massive currents.
In all cases, when the state parameters remain small, only the chiral
massless states are accessible and the regime is obviously subsonic,
as can be seen in Fig.~\ref{figtwomassclct}. However, the massive mode
filling modifies radically this behavior, and as found for the lowest
massive modes alone, as soon as a mass hyperbola is reached, the
longitudinal perturbations propagation speed falls drastically and
ends up being less than the transverse perturbation velocity. There is
a rapid transition from the subsonic to the supersonic regime. For
higher densities $\nu$, the behavior depends on the coupling
constant. More precisely, in the nonperturbative sector, the
ultrarelativistic limit can be applied before the energy scales reach
the fermion vacuum masses, and thus the subsonic regime is recovered,
whereas it is not the case in the perturbative sector, as can be seen
in Fig.~\ref{figtwomassclct}.
\begin{figure}
\begin{center}
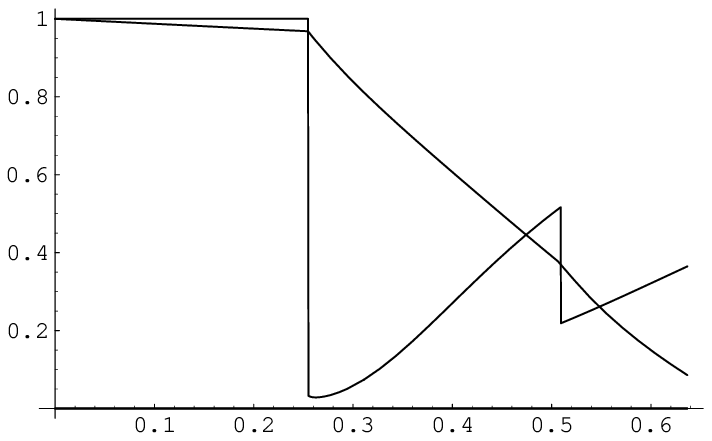
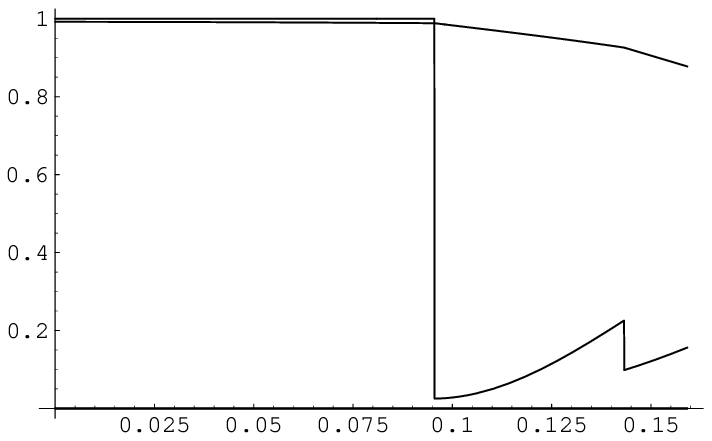
\caption{The squared longitudinal and transverse perturbation
propagation speeds for a spectrum involving two massive states in
addition to the zero mode, plotted as functions of the state parameter
$\nu_\Ferm$ of one species, the other being fixed to a particular
value. The curves have been plotted for two values of the coupling
constant to the Higgs field, $m_{\mathrm{f}}/m_{\mathrm{h}} \sim 1$
and $m_{\mathrm{f}}/m_{\mathrm{h}} \sim 4$. Note the successive
transitions between subsonic and supersonic behaviors according to the
allowed jumps to the mass hyperbolae. However, the fermion vacuum mass
limit does not allow the ultrarelativistic limit to take place in the
perturbative sector, as it was the case for the lowest massive modes
alone. In this case, the string dynamics follows a supersonic regime
as soon as the first massive bound state is filled.}
\label{figtwomassclct}
\end{center}
\end{figure}

\subsubsection{Discussion}

All these results have been derived without considering the back
reaction effects induced by the trapped charge currents along the
string [see Eq.~(\ref{gaugecurrents})]. As was already discussed for
the zero modes in Ref.~\cite{ringeval}, these currents yield back reacted
gauge field, $B_t$ and $B_z$, which might modify the vortex background
and the fermionic equations of motion. However, such perturbations of
the Higgs and orthoradial gauge field profiles (see
Fig.~\ref{figback}) can be neglected for $B_t B^t$ and $B_z B^z$ small
compared to the string forming gauge field $B_\theta B^\theta \sim
m_{\mathrm{b}}^2$. Using Eqs.~(\ref{defalphabar}) and
(\ref{gaugecurrents}), the dimensionless charge currents associated
with one massive bound state, and generating the dimensionless gauge
fields $B_\alpha/m_{\mathrm{b}}$, roughly read
\begin{equation}
\label{roughcurrent}
\widetilde{j}^\alpha \sim \frac{c_\Ferm}{\cphi} \rhot_\Ferm
\frac{\miv_\Ferm}{2 \pi^2 \eta},
\end{equation}
with $c_\Ferm$ the fermion charges, i.e., $\cfr$ or $\cfl$. As a
result, the back reaction on the vortex background is negligible as
long as $\miv < 2 \pi^2 \eta$, which is clearly satisfied in the full
perturbative sector. Moreover, since $m_{\mathrm{h}}=\eta
\sqrt{\lambda}$, the previous derivations of the equation of state are
also valid in the nonperturbative sector provided $\miv < 2 \pi^2
m_{\mathrm{h}}/\sqrt{\lambda}$, and thus depend on the values of
self-coupling constant of the Higgs field $\lambda$, but also on the
mass spectrum. As can be seen in Fig.~\ref{figmasspect}, the ratio
$\miv/m_{\mathrm{f}}$ decreases with the fermion vacuum mass
$m_{\mathrm{f}}$, as a result $\miv/m_{\mathrm{h}}$ increases all the
more slowly, which allows to have both $m_{\mathrm{f}}>m_{\mathrm{h}}$
and $\miv < 2 \pi^2 m_{\mathrm{h}}/\sqrt{\lambda}$.

Moreover, in order for the new gauge fields $B_t$ and $B_z$ to
not significantly modify the fermionic equations of motion, from
Eq.~(\ref{fermmvt}), they have to verify $q \cphi B_\alpha < \omega
\sim \miv_\Ferm$. As a result, Eq.~(\ref{roughcurrent}), and
$B_\alpha/m_{\mathrm{b}} \sim \widetilde{j}^\alpha$ yield the
condition
\begin{equation}
\frac{m_{\mathrm{b}}^2}{\eta^2} \frac{c_\Ferm}{c_\phi} < 1.
\end{equation}
As expected, it is essentially the same condition as the one
previously derived for the zero modes alone~\cite{ringeval}. On the
other hand, although the back reaction on the fermionic equations of
motion can deeply modify the zero mode currents~\cite{ringeval3}, since
the massive bound states are no longer eigenstates of the $\gamma^0
\gamma^3$ operator, it is reasonable to assume that rather than modify
their nature and existence significantly, the back reaction gauge
fields may only modify their mass spectrum. In this sense, back
reaction would indeed be a correction.

\section{Conclusion}

The relevant characteristic features of Dirac fermions trapped in a
cosmic string in the form of massive bound states have been study
numerically, in the framework of the Witten model, and in the neutral
limit. By means of a two-dimensional quantization of the associated
spinor fields along the string world sheet, the energy per unit length
and the tension of a cosmic string carrying any kind of fermionic
current, massive or massless, have been computed, and found to involve
as many state parameters as different trapped modes. However, in the
zero-temperature limit, only two have been found to be relevant and
they can be defined as the density numbers of the chiral zero mode
exitations associated with the two fermions $\Psi$ and $\Chi$ coupled
to the Higgs field.

As a result, it was shown that the fixed trace equation of state no
longer applies, as soon as massive states are filled, i.e., for energy
scales larger than the lowest massive mode belonging to the mass
spectrum. Moreover, the filling of massive states leads to a rapid
transition from the subsonic regime, relevant with massless, or
ultrarelativistic currents, to supersonic. Such properties could be
relevant in vorton evolution since it has been shown that supersonic
regimes generally lead to their classical
instabilities~\cite{cartermartin}. As a result, in the perturbative
sectors for which $m_{\mathrm{f}} < m_{\mathrm{h}}$, the protovortons
could be essentially produced at energy scales necessarily smaller
than the lower mass of the spectrum, where the fermionic currents
consist essentially in zero modes. In this way, vortons with fermionic
currents could be included in the more general two energy scale
models~\cite{brandenberger}. However, the present conclusions are
restricted to parameter domains of the model where the back reaction
can be neglected. Although it is reasonable to consider that the back
reaction effects may simply modify the massive bound states through
their mass values, their influence on zero modes are expected to be
much more significant. In particular, the modified zero modes cannot
be any longer eigenstates of the $\gamma^0 \gamma^3$
operator~\cite{ringeval}, so one may conjecture that they acquire an
effective mass, leading to massive states potentially instable for
cosmic string loops

\acknowledgments It is a pleasure to thank particularly P. Peter for
many fruitful and helpfull discussions. I also wish to acknowledge
M. Lemoine, J. Martin, and O. Poujade for advice and various
enlightening discussions.

\end{document}